\documentclass[longauth]{aaEC}

\usepackage{graphicx}
\usepackage{natbib}
\usepackage{scalerel}
\usepackage{comment}
\usepackage{amsmath}

\usepackage[table]{xcolor}

\usepackage{txfonts}
\usepackage[pdfencoding=auto,psdextra]{hyperref}
\hypersetup{
    colorlinks=true,
    linkcolor=blue,
    filecolor=magenta,      
    urlcolor=blue,
    citecolor=blue
}
\makeatletter
\renewcommand*\aa@pageof{, page \thepage{} of \pageref*{LastPage}}

\makeatother

\usepackage[utf8]{inputenc}

\usepackage[switch, modulo]{lineno}
\renewcommand{\linenumbers}[0]{}

\usepackage{euclid}

\begin{document}
%
%
\title{Euclid Quick Data Release (Q1)}
\subtitle{From simulations to sky: Advancing machine-learning lens detection with real Euclid data}

\newcommand{\orcid}[1]{} 


\author{Euclid Collaboration: N.~E.~P.~Lines\orcid{0009-0004-7751-1914}\thanks{\email{natalie.lines@port.ac.uk}}\inst{\ref{aff1}}
\and T.~E.~Collett\orcid{0000-0001-5564-3140}\inst{\ref{aff1}}
\and P.~Holloway\orcid{0009-0002-8896-6100}\inst{\ref{aff1}}
\and K.~Rojas\orcid{0000-0003-1391-6854}\inst{\ref{aff2}}
\and S.~Schuldt\orcid{0000-0003-2497-6334}\inst{\ref{aff3},\ref{aff4}}
\and R.~B.~Metcalf\orcid{0000-0003-3167-2574}\inst{\ref{aff5},\ref{aff6}}
\and T.~Li\orcid{0009-0005-5008-0381}\inst{\ref{aff1}}
\and A.~Verma\orcid{0000-0002-0730-0781}\inst{\ref{aff7}}
\and G.~Despali\orcid{0000-0001-6150-4112}\inst{\ref{aff5},\ref{aff6},\ref{aff8}}
\and F.~Courbin\orcid{0000-0003-0758-6510}\inst{\ref{aff9},\ref{aff10},\ref{aff11}}
\and R.~Gavazzi\orcid{0000-0002-5540-6935}\inst{\ref{aff12},\ref{aff13}}
\and C.~Tortora\orcid{0000-0001-7958-6531}\inst{\ref{aff14}}
\and B.~Cl\'ement\orcid{0000-0002-7966-3661}\inst{\ref{aff15},\ref{aff16}}
\and N.~Aghanim\orcid{0000-0002-6688-8992}\inst{\ref{aff17}}
\and B.~Altieri\orcid{0000-0003-3936-0284}\inst{\ref{aff18}}
\and L.~Amendola\orcid{0000-0002-0835-233X}\inst{\ref{aff19}}
\and S.~Andreon\orcid{0000-0002-2041-8784}\inst{\ref{aff20}}
\and N.~Auricchio\orcid{0000-0003-4444-8651}\inst{\ref{aff6}}
\and C.~Baccigalupi\orcid{0000-0002-8211-1630}\inst{\ref{aff21},\ref{aff22},\ref{aff23},\ref{aff24}}
\and M.~Baldi\orcid{0000-0003-4145-1943}\inst{\ref{aff25},\ref{aff6},\ref{aff8}}
\and A.~Balestra\orcid{0000-0002-6967-261X}\inst{\ref{aff26}}
\and S.~Bardelli\orcid{0000-0002-8900-0298}\inst{\ref{aff6}}
\and P.~Battaglia\orcid{0000-0002-7337-5909}\inst{\ref{aff6}}
\and A.~Biviano\orcid{0000-0002-0857-0732}\inst{\ref{aff22},\ref{aff21}}
\and E.~Branchini\orcid{0000-0002-0808-6908}\inst{\ref{aff27},\ref{aff28},\ref{aff20}}
\and M.~Brescia\orcid{0000-0001-9506-5680}\inst{\ref{aff29},\ref{aff14}}
\and S.~Camera\orcid{0000-0003-3399-3574}\inst{\ref{aff30},\ref{aff31},\ref{aff32}}
\and G.~Ca\~nas-Herrera\orcid{0000-0003-2796-2149}\inst{\ref{aff33},\ref{aff34}}
\and V.~Capobianco\orcid{0000-0002-3309-7692}\inst{\ref{aff32}}
\and C.~Carbone\orcid{0000-0003-0125-3563}\inst{\ref{aff4}}
\and J.~Carretero\orcid{0000-0002-3130-0204}\inst{\ref{aff35},\ref{aff36}}
\and M.~Castellano\orcid{0000-0001-9875-8263}\inst{\ref{aff37}}
\and G.~Castignani\orcid{0000-0001-6831-0687}\inst{\ref{aff6}}
\and S.~Cavuoti\orcid{0000-0002-3787-4196}\inst{\ref{aff14},\ref{aff38}}
\and A.~Cimatti\inst{\ref{aff39}}
\and C.~Colodro-Conde\inst{\ref{aff40}}
\and G.~Congedo\orcid{0000-0003-2508-0046}\inst{\ref{aff33}}
\and C.~J.~Conselice\orcid{0000-0003-1949-7638}\inst{\ref{aff41}}
\and L.~Conversi\orcid{0000-0002-6710-8476}\inst{\ref{aff42},\ref{aff18}}
\and Y.~Copin\orcid{0000-0002-5317-7518}\inst{\ref{aff43}}
\and H.~M.~Courtois\orcid{0000-0003-0509-1776}\inst{\ref{aff44}}
\and M.~Cropper\orcid{0000-0003-4571-9468}\inst{\ref{aff45}}
\and H.~Degaudenzi\orcid{0000-0002-5887-6799}\inst{\ref{aff46}}
\and G.~De~Lucia\orcid{0000-0002-6220-9104}\inst{\ref{aff22}}
\and H.~Dole\orcid{0000-0002-9767-3839}\inst{\ref{aff17}}
\and F.~Dubath\orcid{0000-0002-6533-2810}\inst{\ref{aff46}}
\and X.~Dupac\inst{\ref{aff18}}
\and S.~Dusini\orcid{0000-0002-1128-0664}\inst{\ref{aff47}}
\and A.~Ealet\orcid{0000-0003-3070-014X}\inst{\ref{aff43}}
\and S.~Escoffier\orcid{0000-0002-2847-7498}\inst{\ref{aff48}}
\and M.~Farina\orcid{0000-0002-3089-7846}\inst{\ref{aff49}}
\and R.~Farinelli\inst{\ref{aff6}}
\and F.~Faustini\orcid{0000-0001-6274-5145}\inst{\ref{aff37},\ref{aff50}}
\and S.~Ferriol\inst{\ref{aff43}}
\and F.~Finelli\orcid{0000-0002-6694-3269}\inst{\ref{aff6},\ref{aff51}}
\and M.~Frailis\orcid{0000-0002-7400-2135}\inst{\ref{aff22}}
\and E.~Franceschi\orcid{0000-0002-0585-6591}\inst{\ref{aff6}}
\and M.~Fumana\orcid{0000-0001-6787-5950}\inst{\ref{aff4}}
\and S.~Galeotta\orcid{0000-0002-3748-5115}\inst{\ref{aff22}}
\and K.~George\orcid{0000-0002-1734-8455}\inst{\ref{aff52}}
\and B.~Gillis\orcid{0000-0002-4478-1270}\inst{\ref{aff33}}
\and C.~Giocoli\orcid{0000-0002-9590-7961}\inst{\ref{aff6},\ref{aff8}}
\and P.~G\'omez-Alvarez\orcid{0000-0002-8594-5358}\inst{\ref{aff53},\ref{aff18}}
\and J.~Gracia-Carpio\inst{\ref{aff54}}
\and A.~Grazian\orcid{0000-0002-5688-0663}\inst{\ref{aff26}}
\and F.~Grupp\inst{\ref{aff54},\ref{aff55}}
\and S.~V.~H.~Haugan\orcid{0000-0001-9648-7260}\inst{\ref{aff56}}
\and W.~Holmes\inst{\ref{aff57}}
\and I.~M.~Hook\orcid{0000-0002-2960-978X}\inst{\ref{aff58}}
\and F.~Hormuth\inst{\ref{aff59}}
\and A.~Hornstrup\orcid{0000-0002-3363-0936}\inst{\ref{aff60},\ref{aff61}}
\and K.~Jahnke\orcid{0000-0003-3804-2137}\inst{\ref{aff62}}
\and M.~Jhabvala\inst{\ref{aff63}}
\and B.~Joachimi\orcid{0000-0001-7494-1303}\inst{\ref{aff64}}
\and E.~Keih\"anen\orcid{0000-0003-1804-7715}\inst{\ref{aff65}}
\and S.~Kermiche\orcid{0000-0002-0302-5735}\inst{\ref{aff48}}
\and A.~Kiessling\orcid{0000-0002-2590-1273}\inst{\ref{aff57}}
\and B.~Kubik\orcid{0009-0006-5823-4880}\inst{\ref{aff43}}
\and M.~K\"ummel\orcid{0000-0003-2791-2117}\inst{\ref{aff55}}
\and M.~Kunz\orcid{0000-0002-3052-7394}\inst{\ref{aff66}}
\and H.~Kurki-Suonio\orcid{0000-0002-4618-3063}\inst{\ref{aff67},\ref{aff68}}
\and A.~M.~C.~Le~Brun\orcid{0000-0002-0936-4594}\inst{\ref{aff69}}
\and S.~Ligori\orcid{0000-0003-4172-4606}\inst{\ref{aff32}}
\and P.~B.~Lilje\orcid{0000-0003-4324-7794}\inst{\ref{aff56}}
\and V.~Lindholm\orcid{0000-0003-2317-5471}\inst{\ref{aff67},\ref{aff68}}
\and I.~Lloro\orcid{0000-0001-5966-1434}\inst{\ref{aff70}}
\and G.~Mainetti\orcid{0000-0003-2384-2377}\inst{\ref{aff71}}
\and D.~Maino\inst{\ref{aff3},\ref{aff4},\ref{aff72}}
\and E.~Maiorano\orcid{0000-0003-2593-4355}\inst{\ref{aff6}}
\and O.~Mansutti\orcid{0000-0001-5758-4658}\inst{\ref{aff22}}
\and S.~Marcin\inst{\ref{aff2}}
\and O.~Marggraf\orcid{0000-0001-7242-3852}\inst{\ref{aff73}}
\and M.~Martinelli\orcid{0000-0002-6943-7732}\inst{\ref{aff37},\ref{aff74}}
\and N.~Martinet\orcid{0000-0003-2786-7790}\inst{\ref{aff12}}
\and F.~Marulli\orcid{0000-0002-8850-0303}\inst{\ref{aff5},\ref{aff6},\ref{aff8}}
\and R.~J.~Massey\orcid{0000-0002-6085-3780}\inst{\ref{aff75}}
\and E.~Medinaceli\orcid{0000-0002-4040-7783}\inst{\ref{aff6}}
\and S.~Mei\orcid{0000-0002-2849-559X}\inst{\ref{aff76},\ref{aff77}}
\and M.~Melchior\inst{\ref{aff78}}
\and Y.~Mellier\inst{\ref{aff79},\ref{aff13}}
\and M.~Meneghetti\orcid{0000-0003-1225-7084}\inst{\ref{aff6},\ref{aff8}}
\and E.~Merlin\orcid{0000-0001-6870-8900}\inst{\ref{aff37}}
\and G.~Meylan\inst{\ref{aff15}}
\and A.~Mora\orcid{0000-0002-1922-8529}\inst{\ref{aff80}}
\and M.~Moresco\orcid{0000-0002-7616-7136}\inst{\ref{aff5},\ref{aff6}}
\and L.~Moscardini\orcid{0000-0002-3473-6716}\inst{\ref{aff5},\ref{aff6},\ref{aff8}}
\and R.~Nakajima\orcid{0009-0009-1213-7040}\inst{\ref{aff73}}
\and C.~Neissner\orcid{0000-0001-8524-4968}\inst{\ref{aff81},\ref{aff36}}
\and S.-M.~Niemi\orcid{0009-0005-0247-0086}\inst{\ref{aff82}}
\and J.~W.~Nightingale\orcid{0000-0002-8987-7401}\inst{\ref{aff83}}
\and C.~Padilla\orcid{0000-0001-7951-0166}\inst{\ref{aff81}}
\and S.~Paltani\orcid{0000-0002-8108-9179}\inst{\ref{aff46}}
\and F.~Pasian\orcid{0000-0002-4869-3227}\inst{\ref{aff22}}
\and K.~Pedersen\inst{\ref{aff84}}
\and W.~J.~Percival\orcid{0000-0002-0644-5727}\inst{\ref{aff85},\ref{aff86},\ref{aff87}}
\and V.~Pettorino\orcid{0000-0002-4203-9320}\inst{\ref{aff82}}
\and S.~Pires\orcid{0000-0002-0249-2104}\inst{\ref{aff88}}
\and G.~Polenta\orcid{0000-0003-4067-9196}\inst{\ref{aff50}}
\and M.~Poncet\inst{\ref{aff89}}
\and L.~A.~Popa\inst{\ref{aff90}}
\and L.~Pozzetti\orcid{0000-0001-7085-0412}\inst{\ref{aff6}}
\and F.~Raison\orcid{0000-0002-7819-6918}\inst{\ref{aff54}}
\and A.~Renzi\orcid{0000-0001-9856-1970}\inst{\ref{aff91},\ref{aff47}}
\and J.~Rhodes\orcid{0000-0002-4485-8549}\inst{\ref{aff57}}
\and G.~Riccio\inst{\ref{aff14}}
\and E.~Romelli\orcid{0000-0003-3069-9222}\inst{\ref{aff22}}
\and M.~Roncarelli\orcid{0000-0001-9587-7822}\inst{\ref{aff6}}
\and C.~Rosset\orcid{0000-0003-0286-2192}\inst{\ref{aff76}}
\and R.~Saglia\orcid{0000-0003-0378-7032}\inst{\ref{aff55},\ref{aff54}}
\and Z.~Sakr\orcid{0000-0002-4823-3757}\inst{\ref{aff19},\ref{aff92},\ref{aff93}}
\and A.~G.~S\'anchez\orcid{0000-0003-1198-831X}\inst{\ref{aff54}}
\and D.~Sapone\orcid{0000-0001-7089-4503}\inst{\ref{aff94}}
\and B.~Sartoris\orcid{0000-0003-1337-5269}\inst{\ref{aff55},\ref{aff22}}
\and J.~A.~Schewtschenko\orcid{0000-0002-4913-6393}\inst{\ref{aff33}}
\and P.~Schneider\orcid{0000-0001-8561-2679}\inst{\ref{aff73}}
\and T.~Schrabback\orcid{0000-0002-6987-7834}\inst{\ref{aff95}}
\and A.~Secroun\orcid{0000-0003-0505-3710}\inst{\ref{aff48}}
\and G.~Seidel\orcid{0000-0003-2907-353X}\inst{\ref{aff62}}
\and S.~Serrano\orcid{0000-0002-0211-2861}\inst{\ref{aff96},\ref{aff97},\ref{aff98}}
\and C.~Sirignano\orcid{0000-0002-0995-7146}\inst{\ref{aff91},\ref{aff47}}
\and G.~Sirri\orcid{0000-0003-2626-2853}\inst{\ref{aff8}}
\and L.~Stanco\orcid{0000-0002-9706-5104}\inst{\ref{aff47}}
\and J.~Steinwagner\orcid{0000-0001-7443-1047}\inst{\ref{aff54}}
\and P.~Tallada-Cresp\'{i}\orcid{0000-0002-1336-8328}\inst{\ref{aff35},\ref{aff36}}
\and A.~N.~Taylor\inst{\ref{aff33}}
\and I.~Tereno\orcid{0000-0002-4537-6218}\inst{\ref{aff99},\ref{aff100}}
\and N.~Tessore\orcid{0000-0002-9696-7931}\inst{\ref{aff45}}
\and S.~Toft\orcid{0000-0003-3631-7176}\inst{\ref{aff101},\ref{aff102}}
\and R.~Toledo-Moreo\orcid{0000-0002-2997-4859}\inst{\ref{aff103}}
\and F.~Torradeflot\orcid{0000-0003-1160-1517}\inst{\ref{aff36},\ref{aff35}}
\and I.~Tutusaus\orcid{0000-0002-3199-0399}\inst{\ref{aff98},\ref{aff96},\ref{aff92}}
\and J.~Valiviita\orcid{0000-0001-6225-3693}\inst{\ref{aff67},\ref{aff68}}
\and T.~Vassallo\orcid{0000-0001-6512-6358}\inst{\ref{aff22}}
\and A.~Veropalumbo\orcid{0000-0003-2387-1194}\inst{\ref{aff20},\ref{aff28},\ref{aff27}}
\and Y.~Wang\orcid{0000-0002-4749-2984}\inst{\ref{aff104}}
\and J.~Weller\orcid{0000-0002-8282-2010}\inst{\ref{aff55},\ref{aff54}}
\and A.~Zacchei\orcid{0000-0003-0396-1192}\inst{\ref{aff22},\ref{aff21}}
\and G.~Zamorani\orcid{0000-0002-2318-301X}\inst{\ref{aff6}}
\and F.~M.~Zerbi\inst{\ref{aff20}}
\and E.~Zucca\orcid{0000-0002-5845-8132}\inst{\ref{aff6}}
\and M.~Ballardini\orcid{0000-0003-4481-3559}\inst{\ref{aff105},\ref{aff106},\ref{aff6}}
\and M.~Bolzonella\orcid{0000-0003-3278-4607}\inst{\ref{aff6}}
\and E.~Bozzo\orcid{0000-0002-8201-1525}\inst{\ref{aff46}}
\and C.~Burigana\orcid{0000-0002-3005-5796}\inst{\ref{aff107},\ref{aff51}}
\and R.~Cabanac\orcid{0000-0001-6679-2600}\inst{\ref{aff92}}
\and M.~Calabrese\orcid{0000-0002-2637-2422}\inst{\ref{aff108},\ref{aff4}}
\and A.~Cappi\inst{\ref{aff109},\ref{aff6}}
\and T.~Castro\orcid{0000-0002-6292-3228}\inst{\ref{aff22},\ref{aff23},\ref{aff21},\ref{aff110}}
\and J.~A.~Escartin~Vigo\inst{\ref{aff54}}
\and L.~Gabarra\orcid{0000-0002-8486-8856}\inst{\ref{aff7}}
\and J.~Garc\'ia-Bellido\orcid{0000-0002-9370-8360}\inst{\ref{aff111}}
\and V.~Gautard\inst{\ref{aff112}}
\and S.~Hemmati\orcid{0000-0003-2226-5395}\inst{\ref{aff104}}
\and M.~Huertas-Company\orcid{0000-0002-1416-8483}\inst{\ref{aff40},\ref{aff113},\ref{aff114}}
\and J.~Macias-Perez\orcid{0000-0002-5385-2763}\inst{\ref{aff115}}
\and R.~Maoli\orcid{0000-0002-6065-3025}\inst{\ref{aff116},\ref{aff37}}
\and J.~Mart\'{i}n-Fleitas\orcid{0000-0002-8594-569X}\inst{\ref{aff117}}
\and M.~Maturi\orcid{0000-0002-3517-2422}\inst{\ref{aff19},\ref{aff118}}
\and N.~Mauri\orcid{0000-0001-8196-1548}\inst{\ref{aff39},\ref{aff8}}
\and P.~Monaco\orcid{0000-0003-2083-7564}\inst{\ref{aff119},\ref{aff22},\ref{aff23},\ref{aff21}}
\and M.~P\"ontinen\orcid{0000-0001-5442-2530}\inst{\ref{aff67}}
\and C.~Porciani\orcid{0000-0002-7797-2508}\inst{\ref{aff73}}
\and I.~Risso\orcid{0000-0003-2525-7761}\inst{\ref{aff20},\ref{aff28}}
\and V.~Scottez\orcid{0009-0008-3864-940X}\inst{\ref{aff79},\ref{aff120}}
\and M.~Sereno\orcid{0000-0003-0302-0325}\inst{\ref{aff6},\ref{aff8}}
\and M.~Tenti\orcid{0000-0002-4254-5901}\inst{\ref{aff8}}
\and M.~Tucci\inst{\ref{aff46}}
\and M.~Viel\orcid{0000-0002-2642-5707}\inst{\ref{aff21},\ref{aff22},\ref{aff24},\ref{aff23},\ref{aff110}}
\and M.~Wiesmann\orcid{0009-0000-8199-5860}\inst{\ref{aff56}}
\and Y.~Akrami\orcid{0000-0002-2407-7956}\inst{\ref{aff111},\ref{aff121}}
\and I.~T.~Andika\orcid{0000-0001-6102-9526}\inst{\ref{aff52},\ref{aff122}}
\and G.~Angora\orcid{0000-0002-0316-6562}\inst{\ref{aff14},\ref{aff105}}
\and S.~Anselmi\orcid{0000-0002-3579-9583}\inst{\ref{aff47},\ref{aff91},\ref{aff123}}
\and M.~Archidiacono\orcid{0000-0003-4952-9012}\inst{\ref{aff3},\ref{aff72}}
\and F.~Atrio-Barandela\orcid{0000-0002-2130-2513}\inst{\ref{aff124}}
\and E.~Aubourg\orcid{0000-0002-5592-023X}\inst{\ref{aff76},\ref{aff125}}
\and L.~Bazzanini\orcid{0000-0003-0727-0137}\inst{\ref{aff105},\ref{aff6}}
\and D.~Bertacca\orcid{0000-0002-2490-7139}\inst{\ref{aff91},\ref{aff26},\ref{aff47}}
\and M.~Bethermin\orcid{0000-0002-3915-2015}\inst{\ref{aff126}}
\and F.~Beutler\orcid{0000-0003-0467-5438}\inst{\ref{aff33}}
\and A.~Blanchard\orcid{0000-0001-8555-9003}\inst{\ref{aff92}}
\and L.~Blot\orcid{0000-0002-9622-7167}\inst{\ref{aff127},\ref{aff69}}
\and M.~Bonici\orcid{0000-0002-8430-126X}\inst{\ref{aff85},\ref{aff4}}
\and S.~Borgani\orcid{0000-0001-6151-6439}\inst{\ref{aff119},\ref{aff21},\ref{aff22},\ref{aff23},\ref{aff110}}
\and M.~L.~Brown\orcid{0000-0002-0370-8077}\inst{\ref{aff41}}
\and S.~Bruton\orcid{0000-0002-6503-5218}\inst{\ref{aff128}}
\and A.~Calabro\orcid{0000-0003-2536-1614}\inst{\ref{aff37}}
\and B.~Camacho~Quevedo\orcid{0000-0002-8789-4232}\inst{\ref{aff21},\ref{aff24},\ref{aff22}}
\and F.~Caro\inst{\ref{aff37}}
\and C.~S.~Carvalho\inst{\ref{aff100}}
\and F.~Cogato\orcid{0000-0003-4632-6113}\inst{\ref{aff5},\ref{aff6}}
\and S.~Conseil\orcid{0000-0002-3657-4191}\inst{\ref{aff43}}
\and A.~R.~Cooray\orcid{0000-0002-3892-0190}\inst{\ref{aff129}}
\and O.~Cucciati\orcid{0000-0002-9336-7551}\inst{\ref{aff6}}
\and S.~Davini\orcid{0000-0003-3269-1718}\inst{\ref{aff28}}
\and F.~De~Paolis\orcid{0000-0001-6460-7563}\inst{\ref{aff130},\ref{aff131},\ref{aff132}}
\and G.~Desprez\orcid{0000-0001-8325-1742}\inst{\ref{aff133}}
\and A.~D\'iaz-S\'anchez\orcid{0000-0003-0748-4768}\inst{\ref{aff134}}
\and S.~Di~Domizio\orcid{0000-0003-2863-5895}\inst{\ref{aff27},\ref{aff28}}
\and J.~M.~Diego\orcid{0000-0001-9065-3926}\inst{\ref{aff135}}
\and P.-A.~Duc\orcid{0000-0003-3343-6284}\inst{\ref{aff126}}
\and V.~Duret\orcid{0009-0009-0383-4960}\inst{\ref{aff48}}
\and M.~Y.~Elkhashab\orcid{0000-0001-9306-2603}\inst{\ref{aff22},\ref{aff23},\ref{aff119},\ref{aff21}}
\and A.~Enia\orcid{0000-0002-0200-2857}\inst{\ref{aff6}}
\and Y.~Fang\orcid{0000-0002-0334-6950}\inst{\ref{aff55}}
\and P.~G.~Ferreira\orcid{0000-0002-3021-2851}\inst{\ref{aff7}}
\and A.~Finoguenov\orcid{0000-0002-4606-5403}\inst{\ref{aff67}}
\and A.~Fontana\orcid{0000-0003-3820-2823}\inst{\ref{aff37}}
\and A.~Franco\orcid{0000-0002-4761-366X}\inst{\ref{aff131},\ref{aff130},\ref{aff132}}
\and K.~Ganga\orcid{0000-0001-8159-8208}\inst{\ref{aff76}}
\and T.~Gasparetto\orcid{0000-0002-7913-4866}\inst{\ref{aff37}}
\and E.~Gaztanaga\orcid{0000-0001-9632-0815}\inst{\ref{aff98},\ref{aff96},\ref{aff1}}
\and F.~Giacomini\orcid{0000-0002-3129-2814}\inst{\ref{aff8}}
\and F.~Gianotti\orcid{0000-0003-4666-119X}\inst{\ref{aff6}}
\and G.~Gozaliasl\orcid{0000-0002-0236-919X}\inst{\ref{aff136},\ref{aff67}}
\and A.~Gruppuso\orcid{0000-0001-9272-5292}\inst{\ref{aff6},\ref{aff8}}
\and M.~Guidi\orcid{0000-0001-9408-1101}\inst{\ref{aff25},\ref{aff6}}
\and C.~M.~Gutierrez\orcid{0000-0001-7854-783X}\inst{\ref{aff137}}
\and A.~Hall\orcid{0000-0002-3139-8651}\inst{\ref{aff33}}
\and H.~Hildebrandt\orcid{0000-0002-9814-3338}\inst{\ref{aff138}}
\and J.~Hjorth\orcid{0000-0002-4571-2306}\inst{\ref{aff84}}
\and J.~J.~E.~Kajava\orcid{0000-0002-3010-8333}\inst{\ref{aff139},\ref{aff140}}
\and Y.~Kang\orcid{0009-0000-8588-7250}\inst{\ref{aff46}}
\and V.~Kansal\orcid{0000-0002-4008-6078}\inst{\ref{aff141},\ref{aff142}}
\and D.~Karagiannis\orcid{0000-0002-4927-0816}\inst{\ref{aff105},\ref{aff143}}
\and K.~Kiiveri\inst{\ref{aff65}}
\and J.~Kim\orcid{0000-0003-2776-2761}\inst{\ref{aff7}}
\and C.~C.~Kirkpatrick\inst{\ref{aff65}}
\and S.~Kruk\orcid{0000-0001-8010-8879}\inst{\ref{aff18}}
\and M.~Lattanzi\orcid{0000-0003-1059-2532}\inst{\ref{aff106}}
\and L.~Legrand\orcid{0000-0003-0610-5252}\inst{\ref{aff144},\ref{aff145}}
\and F.~Lepori\orcid{0009-0000-5061-7138}\inst{\ref{aff146}}
\and G.~Leroy\orcid{0009-0004-2523-4425}\inst{\ref{aff147},\ref{aff75}}
\and G.~F.~Lesci\orcid{0000-0002-4607-2830}\inst{\ref{aff5},\ref{aff6}}
\and J.~Lesgourgues\orcid{0000-0001-7627-353X}\inst{\ref{aff148}}
\and T.~I.~Liaudat\orcid{0000-0002-9104-314X}\inst{\ref{aff125}}
\and M.~Magliocchetti\orcid{0000-0001-9158-4838}\inst{\ref{aff49}}
\and A.~Manj\'on-Garc\'ia\orcid{0000-0002-7413-8825}\inst{\ref{aff134}}
\and F.~Mannucci\orcid{0000-0002-4803-2381}\inst{\ref{aff149}}
\and C.~J.~A.~P.~Martins\orcid{0000-0002-4886-9261}\inst{\ref{aff150},\ref{aff151}}
\and L.~Maurin\orcid{0000-0002-8406-0857}\inst{\ref{aff17}}
\and M.~Miluzio\inst{\ref{aff18},\ref{aff152}}
\and A.~Montoro\orcid{0000-0003-4730-8590}\inst{\ref{aff98},\ref{aff96}}
\and C.~Moretti\orcid{0000-0003-3314-8936}\inst{\ref{aff22},\ref{aff21},\ref{aff23}}
\and G.~Morgante\inst{\ref{aff6}}
\and S.~Nadathur\orcid{0000-0001-9070-3102}\inst{\ref{aff1}}
\and K.~Naidoo\orcid{0000-0002-9182-1802}\inst{\ref{aff1},\ref{aff62}}
\and P.~Natoli\orcid{0000-0003-0126-9100}\inst{\ref{aff105},\ref{aff106}}
\and S.~Nesseris\orcid{0000-0002-0567-0324}\inst{\ref{aff111}}
\and D.~Paoletti\orcid{0000-0003-4761-6147}\inst{\ref{aff6},\ref{aff51}}
\and F.~Passalacqua\orcid{0000-0002-8606-4093}\inst{\ref{aff91},\ref{aff47}}
\and K.~Paterson\orcid{0000-0001-8340-3486}\inst{\ref{aff62}}
\and L.~Patrizii\inst{\ref{aff8}}
\and A.~Pisani\orcid{0000-0002-6146-4437}\inst{\ref{aff48}}
\and D.~Potter\orcid{0000-0002-0757-5195}\inst{\ref{aff146}}
\and G.~W.~Pratt\inst{\ref{aff88}}
\and S.~Quai\orcid{0000-0002-0449-8163}\inst{\ref{aff5},\ref{aff6}}
\and M.~Radovich\orcid{0000-0002-3585-866X}\inst{\ref{aff26}}
\and W.~Roster\orcid{0000-0002-9149-6528}\inst{\ref{aff54}}
\and S.~Sacquegna\orcid{0000-0002-8433-6630}\inst{\ref{aff153}}
\and M.~Sahl\'en\orcid{0000-0003-0973-4804}\inst{\ref{aff154}}
\and D.~B.~Sanders\orcid{0000-0002-1233-9998}\inst{\ref{aff155}}
\and E.~Sarpa\orcid{0000-0002-1256-655X}\inst{\ref{aff24},\ref{aff110},\ref{aff23}}
\and A.~Schneider\orcid{0000-0001-7055-8104}\inst{\ref{aff146}}
\and D.~Sciotti\orcid{0009-0008-4519-2620}\inst{\ref{aff37},\ref{aff74}}
\and E.~Sellentin\inst{\ref{aff156},\ref{aff34}}
\and L.~C.~Smith\orcid{0000-0002-3259-2771}\inst{\ref{aff157}}
\and J.~G.~Sorce\orcid{0000-0002-2307-2432}\inst{\ref{aff158},\ref{aff17}}
\and K.~Tanidis\orcid{0000-0001-9843-5130}\inst{\ref{aff7}}
\and C.~Tao\orcid{0000-0001-7961-8177}\inst{\ref{aff48}}
\and F.~Tarsitano\orcid{0000-0002-5919-0238}\inst{\ref{aff159},\ref{aff46}}
\and G.~Testera\inst{\ref{aff28}}
\and R.~Teyssier\orcid{0000-0001-7689-0933}\inst{\ref{aff160}}
\and S.~Tosi\orcid{0000-0002-7275-9193}\inst{\ref{aff27},\ref{aff28},\ref{aff20}}
\and A.~Troja\orcid{0000-0003-0239-4595}\inst{\ref{aff91},\ref{aff47}}
\and A.~Venhola\orcid{0000-0001-6071-4564}\inst{\ref{aff161}}
\and D.~Vergani\orcid{0000-0003-0898-2216}\inst{\ref{aff6}}
\and G.~Vernardos\orcid{0000-0001-8554-7248}\inst{\ref{aff162},\ref{aff163}}
\and G.~Verza\orcid{0000-0002-1886-8348}\inst{\ref{aff164},\ref{aff165}}
\and S.~Vinciguerra\orcid{0009-0005-4018-3184}\inst{\ref{aff12}}
\and M.~Walmsley\orcid{0000-0002-6408-4181}\inst{\ref{aff166},\ref{aff41}}
\and N.~A.~Walton\orcid{0000-0003-3983-8778}\inst{\ref{aff157}}
\and A.~H.~Wright\orcid{0000-0001-7363-7932}\inst{\ref{aff138}}}
										   
\institute{Institute of Cosmology and Gravitation, University of Portsmouth, Portsmouth PO1 3FX, UK\label{aff1}
\and
University of Applied Sciences and Arts of Northwestern Switzerland, School of Computer Science, 5210 Windisch, Switzerland\label{aff2}
\and
Dipartimento di Fisica "Aldo Pontremoli", Universit\`a degli Studi di Milano, Via Celoria 16, 20133 Milano, Italy\label{aff3}
\and
INAF-IASF Milano, Via Alfonso Corti 12, 20133 Milano, Italy\label{aff4}
\and
Dipartimento di Fisica e Astronomia "Augusto Righi" - Alma Mater Studiorum Universit\`a di Bologna, via Piero Gobetti 93/2, 40129 Bologna, Italy\label{aff5}
\and
INAF-Osservatorio di Astrofisica e Scienza dello Spazio di Bologna, Via Piero Gobetti 93/3, 40129 Bologna, Italy\label{aff6}
\and
Department of Physics, Oxford University, Keble Road, Oxford OX1 3RH, UK\label{aff7}
\and
INFN-Sezione di Bologna, Viale Berti Pichat 6/2, 40127 Bologna, Italy\label{aff8}
\and
Institut de Ci\`{e}ncies del Cosmos (ICCUB), Universitat de Barcelona (IEEC-UB), Mart\'{i} i Franqu\`{e}s 1, 08028 Barcelona, Spain\label{aff9}
\and
Instituci\'o Catalana de Recerca i Estudis Avan\c{c}ats (ICREA), Passeig de Llu\'{\i}s Companys 23, 08010 Barcelona, Spain\label{aff10}
\and
Institut de Ciencies de l'Espai (IEEC-CSIC), Campus UAB, Carrer de Can Magrans, s/n Cerdanyola del Vall\'es, 08193 Barcelona, Spain\label{aff11}
\and
Aix-Marseille Universit\'e, CNRS, CNES, LAM, Marseille, France\label{aff12}
\and
Institut d'Astrophysique de Paris, UMR 7095, CNRS, and Sorbonne Universit\'e, 98 bis boulevard Arago, 75014 Paris, France\label{aff13}
\and
INAF-Osservatorio Astronomico di Capodimonte, Via Moiariello 16, 80131 Napoli, Italy\label{aff14}
\and
Institute of Physics, Laboratory of Astrophysics, Ecole Polytechnique F\'ed\'erale de Lausanne (EPFL), Observatoire de Sauverny, 1290 Versoix, Switzerland\label{aff15}
\and
SCITAS, Ecole Polytechnique F\'ed\'erale de Lausanne (EPFL), 1015 Lausanne, Switzerland\label{aff16}
\and
Universit\'e Paris-Saclay, CNRS, Institut d'astrophysique spatiale, 91405, Orsay, France\label{aff17}
\and
ESAC/ESA, Camino Bajo del Castillo, s/n., Urb. Villafranca del Castillo, 28692 Villanueva de la Ca\~nada, Madrid, Spain\label{aff18}
\and
Institut f\"ur Theoretische Physik, University of Heidelberg, Philosophenweg 16, 69120 Heidelberg, Germany\label{aff19}
\and
INAF-Osservatorio Astronomico di Brera, Via Brera 28, 20122 Milano, Italy\label{aff20}
\and
IFPU, Institute for Fundamental Physics of the Universe, via Beirut 2, 34151 Trieste, Italy\label{aff21}
\and
INAF-Osservatorio Astronomico di Trieste, Via G. B. Tiepolo 11, 34143 Trieste, Italy\label{aff22}
\and
INFN, Sezione di Trieste, Via Valerio 2, 34127 Trieste TS, Italy\label{aff23}
\and
SISSA, International School for Advanced Studies, Via Bonomea 265, 34136 Trieste TS, Italy\label{aff24}
\and
Dipartimento di Fisica e Astronomia, Universit\`a di Bologna, Via Gobetti 93/2, 40129 Bologna, Italy\label{aff25}
\and
INAF-Osservatorio Astronomico di Padova, Via dell'Osservatorio 5, 35122 Padova, Italy\label{aff26}
\and
Dipartimento di Fisica, Universit\`a di Genova, Via Dodecaneso 33, 16146, Genova, Italy\label{aff27}
\and
INFN-Sezione di Genova, Via Dodecaneso 33, 16146, Genova, Italy\label{aff28}
\and
Department of Physics "E. Pancini", University Federico II, Via Cinthia 6, 80126, Napoli, Italy\label{aff29}
\and
Dipartimento di Fisica, Universit\`a degli Studi di Torino, Via P. Giuria 1, 10125 Torino, Italy\label{aff30}
\and
INFN-Sezione di Torino, Via P. Giuria 1, 10125 Torino, Italy\label{aff31}
\and
INAF-Osservatorio Astrofisico di Torino, Via Osservatorio 20, 10025 Pino Torinese (TO), Italy\label{aff32}
\and
Institute for Astronomy, University of Edinburgh, Royal Observatory, Blackford Hill, Edinburgh EH9 3HJ, UK\label{aff33}
\and
Leiden Observatory, Leiden University, Einsteinweg 55, 2333 CC Leiden, The Netherlands\label{aff34}
\and
Centro de Investigaciones Energ\'eticas, Medioambientales y Tecnol\'ogicas (CIEMAT), Avenida Complutense 40, 28040 Madrid, Spain\label{aff35}
\and
Port d'Informaci\'{o} Cient\'{i}fica, Campus UAB, C. Albareda s/n, 08193 Bellaterra (Barcelona), Spain\label{aff36}
\and
INAF-Osservatorio Astronomico di Roma, Via Frascati 33, 00078 Monteporzio Catone, Italy\label{aff37}
\and
INFN section of Naples, Via Cinthia 6, 80126, Napoli, Italy\label{aff38}
\and
Dipartimento di Fisica e Astronomia "Augusto Righi" - Alma Mater Studiorum Universit\`a di Bologna, Viale Berti Pichat 6/2, 40127 Bologna, Italy\label{aff39}
\and
Instituto de Astrof\'{\i}sica de Canarias, E-38205 La Laguna, Tenerife, Spain\label{aff40}
\and
Jodrell Bank Centre for Astrophysics, Department of Physics and Astronomy, University of Manchester, Oxford Road, Manchester M13 9PL, UK\label{aff41}
\and
European Space Agency/ESRIN, Largo Galileo Galilei 1, 00044 Frascati, Roma, Italy\label{aff42}
\and
Universit\'e Claude Bernard Lyon 1, CNRS/IN2P3, IP2I Lyon, UMR 5822, Villeurbanne, F-69100, France\label{aff43}
\and
UCB Lyon 1, CNRS/IN2P3, IUF, IP2I Lyon, 4 rue Enrico Fermi, 69622 Villeurbanne, France\label{aff44}
\and
Mullard Space Science Laboratory, University College London, Holmbury St Mary, Dorking, Surrey RH5 6NT, UK\label{aff45}
\and
Department of Astronomy, University of Geneva, ch. d'Ecogia 16, 1290 Versoix, Switzerland\label{aff46}
\and
INFN-Padova, Via Marzolo 8, 35131 Padova, Italy\label{aff47}
\and
Aix-Marseille Universit\'e, CNRS/IN2P3, CPPM, Marseille, France\label{aff48}
\and
INAF-Istituto di Astrofisica e Planetologia Spaziali, via del Fosso del Cavaliere, 100, 00100 Roma, Italy\label{aff49}
\and
Space Science Data Center, Italian Space Agency, via del Politecnico snc, 00133 Roma, Italy\label{aff50}
\and
INFN-Bologna, Via Irnerio 46, 40126 Bologna, Italy\label{aff51}
\and
University Observatory, LMU Faculty of Physics, Scheinerstr.~1, 81679 Munich, Germany\label{aff52}
\and
FRACTAL S.L.N.E., calle Tulip\'an 2, Portal 13 1A, 28231, Las Rozas de Madrid, Spain\label{aff53}
\and
Max Planck Institute for Extraterrestrial Physics, Giessenbachstr. 1, 85748 Garching, Germany\label{aff54}
\and
Universit\"ats-Sternwarte M\"unchen, Fakult\"at f\"ur Physik, Ludwig-Maximilians-Universit\"at M\"unchen, Scheinerstr.~1, 81679 M\"unchen, Germany\label{aff55}
\and
Institute of Theoretical Astrophysics, University of Oslo, P.O. Box 1029 Blindern, 0315 Oslo, Norway\label{aff56}
\and
Jet Propulsion Laboratory, California Institute of Technology, 4800 Oak Grove Drive, Pasadena, CA, 91109, USA\label{aff57}
\and
Department of Physics, Lancaster University, Lancaster, LA1 4YB, UK\label{aff58}
\and
Felix Hormuth Engineering, Goethestr. 17, 69181 Leimen, Germany\label{aff59}
\and
Technical University of Denmark, Elektrovej 327, 2800 Kgs. Lyngby, Denmark\label{aff60}
\and
Cosmic Dawn Center (DAWN), Denmark\label{aff61}
\and
Max-Planck-Institut f\"ur Astronomie, K\"onigstuhl 17, 69117 Heidelberg, Germany\label{aff62}
\and
NASA Goddard Space Flight Center, Greenbelt, MD 20771, USA\label{aff63}
\and
Department of Physics and Astronomy, University College London, Gower Street, London WC1E 6BT, UK\label{aff64}
\and
Department of Physics and Helsinki Institute of Physics, Gustaf H\"allstr\"omin katu 2, University of Helsinki, 00014 Helsinki, Finland\label{aff65}
\and
Universit\'e de Gen\`eve, D\'epartement de Physique Th\'eorique and Centre for Astroparticle Physics, 24 quai Ernest-Ansermet, CH-1211 Gen\`eve 4, Switzerland\label{aff66}
\and
Department of Physics, P.O. Box 64, University of Helsinki, 00014 Helsinki, Finland\label{aff67}
\and
Helsinki Institute of Physics, Gustaf H{\"a}llstr{\"o}min katu 2, University of Helsinki, 00014 Helsinki, Finland\label{aff68}
\and
Laboratoire d'etude de l'Univers et des phenomenes eXtremes, Observatoire de Paris, Universit\'e PSL, Sorbonne Universit\'e, CNRS, 92190 Meudon, France\label{aff69}
\and
SKAO, Jodrell Bank, Lower Withington, Macclesfield SK11 9FT, UK\label{aff70}
\and
Centre de Calcul de l'IN2P3/CNRS, 21 avenue Pierre de Coubertin 69627 Villeurbanne Cedex, France\label{aff71}
\and
INFN-Sezione di Milano, Via Celoria 16, 20133 Milano, Italy\label{aff72}
\and
Universit\"at Bonn, Argelander-Institut f\"ur Astronomie, Auf dem H\"ugel 71, 53121 Bonn, Germany\label{aff73}
\and
INFN-Sezione di Roma, Piazzale Aldo Moro, 2 - c/o Dipartimento di Fisica, Edificio G. Marconi, 00185 Roma, Italy\label{aff74}
\and
Department of Physics, Institute for Computational Cosmology, Durham University, South Road, Durham, DH1 3LE, UK\label{aff75}
\and
Universit\'e Paris Cit\'e, CNRS, Astroparticule et Cosmologie, 75013 Paris, France\label{aff76}
\and
CNRS-UCB International Research Laboratory, Centre Pierre Bin\'etruy, IRL2007, CPB-IN2P3, Berkeley, USA\label{aff77}
\and
University of Applied Sciences and Arts of Northwestern Switzerland, School of Engineering, 5210 Windisch, Switzerland\label{aff78}
\and
Institut d'Astrophysique de Paris, 98bis Boulevard Arago, 75014, Paris, France\label{aff79}
\and
Telespazio UK S.L. for European Space Agency (ESA), Camino bajo del Castillo, s/n, Urbanizacion Villafranca del Castillo, Villanueva de la Ca\~nada, 28692 Madrid, Spain\label{aff80}
\and
Institut de F\'{i}sica d'Altes Energies (IFAE), The Barcelona Institute of Science and Technology, Campus UAB, 08193 Bellaterra (Barcelona), Spain\label{aff81}
\and
European Space Agency/ESTEC, Keplerlaan 1, 2201 AZ Noordwijk, The Netherlands\label{aff82}
\and
School of Mathematics, Statistics and Physics, Newcastle University, Herschel Building, Newcastle-upon-Tyne, NE1 7RU, UK\label{aff83}
\and
DARK, Niels Bohr Institute, University of Copenhagen, Jagtvej 155, 2200 Copenhagen, Denmark\label{aff84}
\and
Waterloo Centre for Astrophysics, University of Waterloo, Waterloo, Ontario N2L 3G1, Canada\label{aff85}
\and
Department of Physics and Astronomy, University of Waterloo, Waterloo, Ontario N2L 3G1, Canada\label{aff86}
\and
Perimeter Institute for Theoretical Physics, Waterloo, Ontario N2L 2Y5, Canada\label{aff87}
\and
Universit\'e Paris-Saclay, Universit\'e Paris Cit\'e, CEA, CNRS, AIM, 91191, Gif-sur-Yvette, France\label{aff88}
\and
Centre National d'Etudes Spatiales -- Centre spatial de Toulouse, 18 avenue Edouard Belin, 31401 Toulouse Cedex 9, France\label{aff89}
\and
Institute of Space Science, Str. Atomistilor, nr. 409 M\u{a}gurele, Ilfov, 077125, Romania\label{aff90}
\and
Dipartimento di Fisica e Astronomia "G. Galilei", Universit\`a di Padova, Via Marzolo 8, 35131 Padova, Italy\label{aff91}
\and
Institut de Recherche en Astrophysique et Plan\'etologie (IRAP), Universit\'e de Toulouse, CNRS, UPS, CNES, 14 Av. Edouard Belin, 31400 Toulouse, France\label{aff92}
\and
Universit\'e St Joseph; Faculty of Sciences, Beirut, Lebanon\label{aff93}
\and
Departamento de F\'isica, FCFM, Universidad de Chile, Blanco Encalada 2008, Santiago, Chile\label{aff94}
\and
Universit\"at Innsbruck, Institut f\"ur Astro- und Teilchenphysik, Technikerstr. 25/8, 6020 Innsbruck, Austria\label{aff95}
\and
Institut d'Estudis Espacials de Catalunya (IEEC),  Edifici RDIT, Campus UPC, 08860 Castelldefels, Barcelona, Spain\label{aff96}
\and
Satlantis, University Science Park, Sede Bld 48940, Leioa-Bilbao, Spain\label{aff97}
\and
Institute of Space Sciences (ICE, CSIC), Campus UAB, Carrer de Can Magrans, s/n, 08193 Barcelona, Spain\label{aff98}
\and
Departamento de F\'isica, Faculdade de Ci\^encias, Universidade de Lisboa, Edif\'icio C8, Campo Grande, PT1749-016 Lisboa, Portugal\label{aff99}
\and
Instituto de Astrof\'isica e Ci\^encias do Espa\c{c}o, Faculdade de Ci\^encias, Universidade de Lisboa, Tapada da Ajuda, 1349-018 Lisboa, Portugal\label{aff100}
\and
Cosmic Dawn Center (DAWN)\label{aff101}
\and
Niels Bohr Institute, University of Copenhagen, Jagtvej 128, 2200 Copenhagen, Denmark\label{aff102}
\and
Universidad Polit\'ecnica de Cartagena, Departamento de Electr\'onica y Tecnolog\'ia de Computadoras,  Plaza del Hospital 1, 30202 Cartagena, Spain\label{aff103}
\and
Caltech/IPAC, 1200 E. California Blvd., Pasadena, CA 91125, USA\label{aff104}
\and
Dipartimento di Fisica e Scienze della Terra, Universit\`a degli Studi di Ferrara, Via Giuseppe Saragat 1, 44122 Ferrara, Italy\label{aff105}
\and
Istituto Nazionale di Fisica Nucleare, Sezione di Ferrara, Via Giuseppe Saragat 1, 44122 Ferrara, Italy\label{aff106}
\and
INAF, Istituto di Radioastronomia, Via Piero Gobetti 101, 40129 Bologna, Italy\label{aff107}
\and
Astronomical Observatory of the Autonomous Region of the Aosta Valley (OAVdA), Loc. Lignan 39, I-11020, Nus (Aosta Valley), Italy\label{aff108}
\and
Universit\'e C\^{o}te d'Azur, Observatoire de la C\^{o}te d'Azur, CNRS, Laboratoire Lagrange, Bd de l'Observatoire, CS 34229, 06304 Nice cedex 4, France\label{aff109}
\and
ICSC - Centro Nazionale di Ricerca in High Performance Computing, Big Data e Quantum Computing, Via Magnanelli 2, Bologna, Italy\label{aff110}
\and
Instituto de F\'isica Te\'orica UAM-CSIC, Campus de Cantoblanco, 28049 Madrid, Spain\label{aff111}
\and
CEA Saclay, DFR/IRFU, Service d'Astrophysique, Bat. 709, 91191 Gif-sur-Yvette, France\label{aff112}
\and
Universit\'e PSL, Observatoire de Paris, Sorbonne Universit\'e, CNRS, LERMA, 75014, Paris, France\label{aff113}
\and
Universit\'e Paris-Cit\'e, 5 Rue Thomas Mann, 75013, Paris, France\label{aff114}
\and
Univ. Grenoble Alpes, CNRS, Grenoble INP, LPSC-IN2P3, 53, Avenue des Martyrs, 38000, Grenoble, France\label{aff115}
\and
Dipartimento di Fisica, Sapienza Universit\`a di Roma, Piazzale Aldo Moro 2, 00185 Roma, Italy\label{aff116}
\and
Aurora Technology for European Space Agency (ESA), Camino bajo del Castillo, s/n, Urbanizacion Villafranca del Castillo, Villanueva de la Ca\~nada, 28692 Madrid, Spain\label{aff117}
\and
Zentrum f\"ur Astronomie, Universit\"at Heidelberg, Philosophenweg 12, 69120 Heidelberg, Germany\label{aff118}
\and
Dipartimento di Fisica - Sezione di Astronomia, Universit\`a di Trieste, Via Tiepolo 11, 34131 Trieste, Italy\label{aff119}
\and
ICL, Junia, Universit\'e Catholique de Lille, LITL, 59000 Lille, France\label{aff120}
\and
CERCA/ISO, Department of Physics, Case Western Reserve University, 10900 Euclid Avenue, Cleveland, OH 44106, USA\label{aff121}
\and
Technical University of Munich, TUM School of Natural Sciences, Physics Department, James-Franck-Str.~1, 85748 Garching, Germany\label{aff122}
\and
Laboratoire Univers et Th\'eorie, Observatoire de Paris, Universit\'e PSL, Universit\'e Paris Cit\'e, CNRS, 92190 Meudon, France\label{aff123}
\and
Departamento de F{\'\i}sica Fundamental. Universidad de Salamanca. Plaza de la Merced s/n. 37008 Salamanca, Spain\label{aff124}
\and
IRFU, CEA, Universit\'e Paris-Saclay 91191 Gif-sur-Yvette Cedex, France\label{aff125}
\and
Universit\'e de Strasbourg, CNRS, Observatoire astronomique de Strasbourg, UMR 7550, 67000 Strasbourg, France\label{aff126}
\and
Center for Data-Driven Discovery, Kavli IPMU (WPI), UTIAS, The University of Tokyo, Kashiwa, Chiba 277-8583, Japan\label{aff127}
\and
California Institute of Technology, 1200 E California Blvd, Pasadena, CA 91125, USA\label{aff128}
\and
Department of Physics \& Astronomy, University of California Irvine, Irvine CA 92697, USA\label{aff129}
\and
Department of Mathematics and Physics E. De Giorgi, University of Salento, Via per Arnesano, CP-I93, 73100, Lecce, Italy\label{aff130}
\and
INFN, Sezione di Lecce, Via per Arnesano, CP-193, 73100, Lecce, Italy\label{aff131}
\and
INAF-Sezione di Lecce, c/o Dipartimento Matematica e Fisica, Via per Arnesano, 73100, Lecce, Italy\label{aff132}
\and
Kapteyn Astronomical Institute, University of Groningen, PO Box 800, 9700 AV Groningen, The Netherlands\label{aff133}
\and
Departamento F\'isica Aplicada, Universidad Polit\'ecnica de Cartagena, Campus Muralla del Mar, 30202 Cartagena, Murcia, Spain\label{aff134}
\and
Instituto de F\'isica de Cantabria, Edificio Juan Jord\'a, Avenida de los Castros, 39005 Santander, Spain\label{aff135}
\and
Department of Computer Science, Aalto University, PO Box 15400, Espoo, FI-00 076, Finland\label{aff136}
\and
 Instituto de Astrof\'{\i}sica de Canarias, E-38205 La Laguna; Universidad de La Laguna, Dpto. Astrof\'\i sica, E-38206 La Laguna, Tenerife, Spain\label{aff137}
\and
Ruhr University Bochum, Faculty of Physics and Astronomy, Astronomical Institute (AIRUB), German Centre for Cosmological Lensing (GCCL), 44780 Bochum, Germany\label{aff138}
\and
Department of Physics and Astronomy, Vesilinnantie 5, University of Turku, 20014 Turku, Finland\label{aff139}
\and
Serco for European Space Agency (ESA), Camino bajo del Castillo, s/n, Urbanizacion Villafranca del Castillo, Villanueva de la Ca\~nada, 28692 Madrid, Spain\label{aff140}
\and
ARC Centre of Excellence for Dark Matter Particle Physics, Melbourne, Australia\label{aff141}
\and
Centre for Astrophysics \& Supercomputing, Swinburne University of Technology,  Hawthorn, Victoria 3122, Australia\label{aff142}
\and
Department of Physics and Astronomy, University of the Western Cape, Bellville, Cape Town, 7535, South Africa\label{aff143}
\and
DAMTP, Centre for Mathematical Sciences, Wilberforce Road, Cambridge CB3 0WA, UK\label{aff144}
\and
Kavli Institute for Cosmology Cambridge, Madingley Road, Cambridge, CB3 0HA, UK\label{aff145}
\and
Department of Astrophysics, University of Zurich, Winterthurerstrasse 190, 8057 Zurich, Switzerland\label{aff146}
\and
Department of Physics, Centre for Extragalactic Astronomy, Durham University, South Road, Durham, DH1 3LE, UK\label{aff147}
\and
Institute for Theoretical Particle Physics and Cosmology (TTK), RWTH Aachen University, 52056 Aachen, Germany\label{aff148}
\and
INAF-Osservatorio Astrofisico di Arcetri, Largo E. Fermi 5, 50125, Firenze, Italy\label{aff149}
\and
Centro de Astrof\'{\i}sica da Universidade do Porto, Rua das Estrelas, 4150-762 Porto, Portugal\label{aff150}
\and
Instituto de Astrof\'isica e Ci\^encias do Espa\c{c}o, Universidade do Porto, CAUP, Rua das Estrelas, PT4150-762 Porto, Portugal\label{aff151}
\and
HE Space for European Space Agency (ESA), Camino bajo del Castillo, s/n, Urbanizacion Villafranca del Castillo, Villanueva de la Ca\~nada, 28692 Madrid, Spain\label{aff152}
\and
INAF - Osservatorio Astronomico d'Abruzzo, Via Maggini, 64100, Teramo, Italy\label{aff153}
\and
Theoretical astrophysics, Department of Physics and Astronomy, Uppsala University, Box 516, 751 37 Uppsala, Sweden\label{aff154}
\and
Institute for Astronomy, University of Hawaii, 2680 Woodlawn Drive, Honolulu, HI 96822, USA\label{aff155}
\and
Mathematical Institute, University of Leiden, Einsteinweg 55, 2333 CA Leiden, The Netherlands\label{aff156}
\and
Institute of Astronomy, University of Cambridge, Madingley Road, Cambridge CB3 0HA, UK\label{aff157}
\and
Univ. Lille, CNRS, Centrale Lille, UMR 9189 CRIStAL, 59000 Lille, France\label{aff158}
\and
Institute for Particle Physics and Astrophysics, Dept. of Physics, ETH Zurich, Wolfgang-Pauli-Strasse 27, 8093 Zurich, Switzerland\label{aff159}
\and
Department of Astrophysical Sciences, Peyton Hall, Princeton University, Princeton, NJ 08544, USA\label{aff160}
\and
Space physics and astronomy research unit, University of Oulu, Pentti Kaiteran katu 1, FI-90014 Oulu, Finland\label{aff161}
\and
Department of Physics and Astronomy, Lehman College of the CUNY, Bronx, NY 10468, USA\label{aff162}
\and
American Museum of Natural History, Department of Astrophysics, New York, NY 10024, USA\label{aff163}
\and
International Centre for Theoretical Physics (ICTP), Strada Costiera 11, 34151 Trieste, Italy\label{aff164}
\and
Center for Computational Astrophysics, Flatiron Institute, 162 5th Avenue, 10010, New York, NY, USA\label{aff165}
\and
David A. Dunlap Department of Astronomy \& Astrophysics, University of Toronto, 50 St George Street, Toronto, Ontario M5S 3H4, Canada\label{aff166}}

%
%

\abstract{
In the era of large-scale surveys like \Euclid, machine learning has become an essential tool for identifying rare yet scientifically valuable objects, such as strong gravitational lenses.
However, supervised machine-learning approaches require large quantities of labelled examples to train on, and the limited number of known strong lenses has lead to a reliance on simulations for training.
A well-known challenge is that machine-learning models trained on one data domain often underperform when applied to a different domain: in the context of lens finding, this means that strong performance on simulated lenses does not necessarily translate into equally good performance on real observations.
In \textcolor{black}{\Euclid's Quick Data Release 1 (Q1)}, covering 63\,deg$^{2}$, 500 strong lens candidates were discovered through a synergy of machine learning, citizen science, and expert visual inspection.
These discoveries now allow us to quantify this performance gap and investigate the impact of training on real data. We find that a network trained only on simulations recovers up to 92\% of simulated lenses with 100\% purity, but only achieves 50\% completeness with 24\% purity on real \Euclid data. By augmenting training data with real \Euclid lenses and non-lenses, completeness improves by 25--30\% in terms of the expected yield of discoverable lenses in \Euclid's Data Release 1 and the full Euclid Wide Survey. 
Roughly 20\% of this improvement comes from the inclusion of real lenses in the training data, while 5--10\% comes from exposure to a more diverse set of non-lenses and false-positives from Q1.
We show that the most effective lens-finding strategy for real-world performance combines the diversity of simulations with the fidelity of real lenses. This hybrid approach establishes a clear methodology for maximising lens discoveries in future data releases from \Euclid, and will likely also be applicable to other surveys such as the Vera Rubin Observatory's Legacy Survey of Space and Time.}

%
%
%
%
   \titlerunning{Euclid Q1: From simulations to sky}
   \authorrunning{Euclid Collaboration: N. E. P. Lines et al.}
   
   \maketitle
%
%
%
%
   
\section{\label{sc:Intro}Introduction}

Strong gravitational lensing is a unique and powerful probe of astrophysics and cosmology, providing direct insights into phenomena that are otherwise difficult to observe. By mapping the deflection of light, strong lenses allow a direct measurement of the total mass of the lens, tracing the combination of its luminous and dark matter. 
\textcolor{black}{On the galaxy-scale, this provides a method of probing the total dark matter contributions \citep{paper1, paper2} and the presence of dark matter subhaloes (e.g., \citealt{vegetti2010, vegetti2012, conor_euclid_substrc, ertl24}), as well as other mass components such as supermassive black holes \citep{cosmichorseshoe,Nightingale2023}.}
\textcolor{black}{Cluster-scale strong lenses can additionally provide insights into dark matter on small scales (e.g., \citealt{nata2017, mene2020, mene2022, mene2023, dutra25}) and offer exceptional magnification and resolving power for the study of distant background sources (e.g., \citealt{vanz17, vanz2020, vanz23, adamo24, mestric22, welch22, grapes, bradley24}). 
Time-delay measurements from strongly lensed transients on both galaxy and cluster-scales provide an independent route to constrain the Hubble constant (e.g., \citealt{kelly23, grillo24, tdcosmo25, snhope, sherry25}), and strong lensing cosmography can provide measurements of cosmological parameters such as the equation of state of dark energy (e.g., \citealt{jullo2010,caminha16, cam22, moresco22, Tian2024}).}
Strong lenses are scientifically valuable, but unfortunately are intrinsically rare, with only a few thousand candidate systems known.

\Euclid \citep{EuclidSkyOverview} is set to revolutionise strong lensing through its unique synergy of wide-field coverage ($\num{14000}$ deg$^2$) and high angular resolution (\ang{;;0.16} \textcolor{black}{point-spread-function (PSF) full-width-half-max} in the optical filter) across the Euclid Wide Survey (EWS).
Forecasts predict that around \num{170000} galaxy-scale strong lenses should be detectable in the full survey \citep{collett15, AcevedoBarroso24}. This unprecedented sample, exceeding the total number of previously known strong lenses by around two orders of magnitude, will be transformative for the field. This vast data set will enable a wealth of new scientific insights into galaxy evolution, dark matter, and cosmology.

With these unprecedented quantities of data comes a new era of lens searching.
Currently, the most reliable method for finding strong lenses is visual inspection by experts \citep{Pearce-Casey24, AcevedoBarroso24}.
However, visually inspecting all 1.5 billion galaxies imaged by \Euclid in order to uncover these strong lenses is intractable, and hence help from automated techniques is necessary. 
Because of the complexity of the lens-finding challenge, deep machine-learning (ML) networks have proved to be one of the most promising methods for addressing it.
In recent years, ML models have been successfully used to find strong lenses in a wealth of astronomical data. These span ground-based large-area surveys such as the Canada--France--Hawaii Telescope Legacy Survey (CFHTLS; e.g., \citealt{Jacobs17}), Dark Energy Survey (DES; e.g., \citealt{Jacobs2018, Jacobs2019,rojas22, jimena}),
Hyper Suprime-Cam (HSC; e.g., \citealt{sonnenfeldSurveyGravitationallylensedObjects2018,sonnenfeldSurveyGravitationallylensedObjects2020a,Canameras2021,shu22_HOLISMOKES8, wongSurveyGravitationallyLensed2022,jaelani24,    schuldt_holismokes_2025, schuldtEtAl25b}),
Kilo-Degree Survey (KiDS; e.g., \citealt{petrillo2017,petrillo--19,li2020,li2021,nagam23,nagam24,Grespan24}), 
Ultraviolet Near-Infrared Optical Northern Survey (UNIONS; e.g., \citealt{savaryStrongLensingUNIONS2022,edgeons2}),
Dark Energy Spectroscopic Instrument (DESI) Legacy Survey (e.g., \citealt{desihuang2020,desihuang2021,desistorfer2024}), and 
Panoramic Survey Telescope and Rapid Response System (PanSTARRS; e.g., \citealt{Canameras20}), as well as space-based imaging from the \HST (HST; e.g., \citealt{hstlensflow, anotherhstpaper}).

The first major data release from the \Euclid survey, Quick Release 1 (Q1; \citealt[]{Q1-TP001}), covers 63\,deg$^2$, and provides the first glimpse into what \Euclid can achieve. In the initial Q1 search, around 500 galaxy-scale strong lens candidates were found through a combination of ML, citizen science, and expert visual inspection, covered by the paper series `The Strong Lensing Discovery Engine’ (SLDE), namely \citet{Q1-SP048}, \citet{Q1-SP052}, \citet{Q1-SP053}, \citet{Q1-SP054}, and \citet{Q1-SP059}.
\textcolor{black}{In addition, around 80 group and cluster-scale lenses from the visual inspection of cluster fields \citep{Q1-SP057}.}
Compared to future releases, a relatively large proportion of Q1 data could be visually inspected to enable this quantity of lens discoveries. However, as the data volume grows, the inherent limits of visual inspection necessitate complementary advances in ML techniques to keep up with the growing volumes of data.

\begin{figure}
    \centering
    \includegraphics[width=1\linewidth]{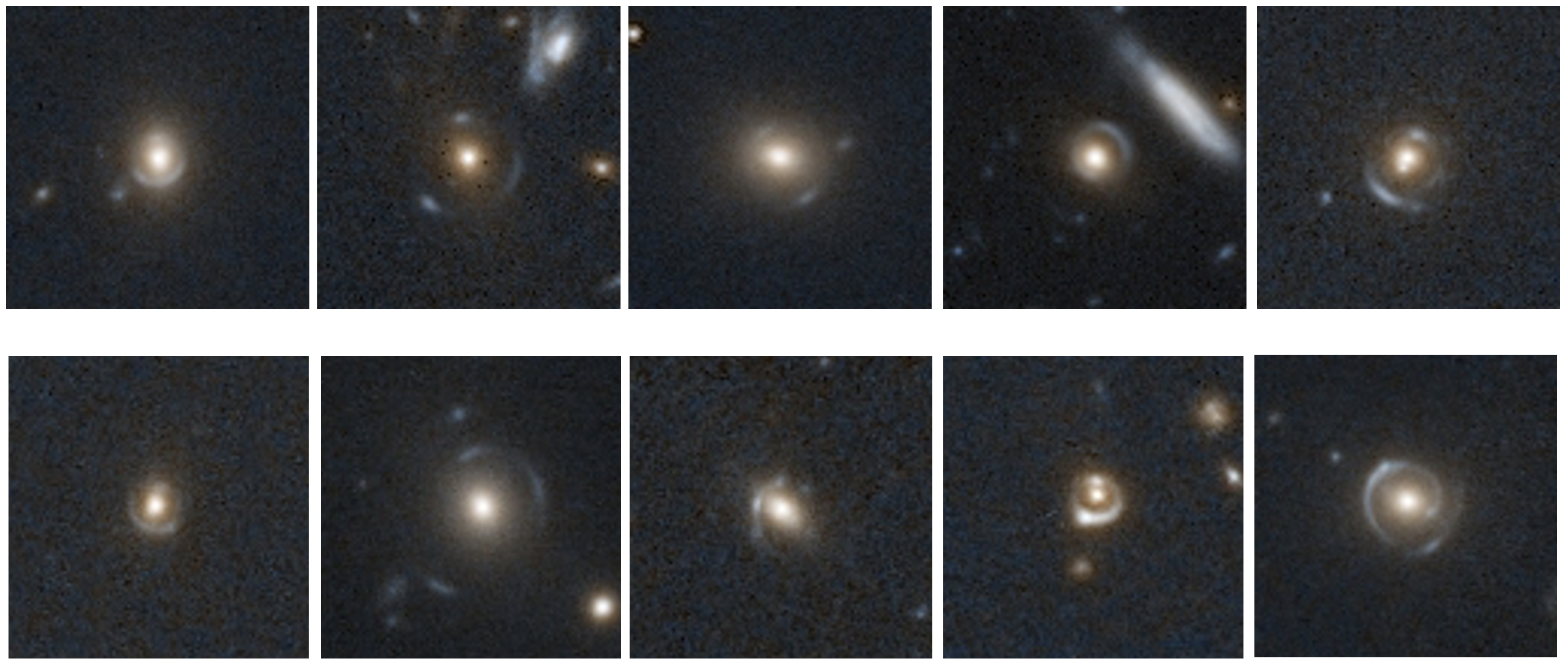}
    \caption{\textit{Top}: simulated \Euclid lenses from \citet{Q1-SP052}. \textit{Bottom}: real lenses found in Q1.}
    \label{fig:simims}
\end{figure}

ML algorithms are natural interpolators: trained on a fixed data set, they typically perform well on unseen data from the same underlying distribution.
However, the known strong lenses are few and far between, and this scarcity poses a problem for training deep-learning models.
Training a supervised ML model from scratch typically requires at least $10^4$--$10^5$, but ideally millions, of labelled images to train on \citep{sun2017}. The number of identified lens candidates within any one survey is typically of the order $10^2$--$10^3$ at most, and hence training ML models for detecting lenses generally relies on simulations.
Although these simulations are made to be as realistic as possible by adding lensed arcs on top of real images of lensing galaxies, resulting in simulations that appear visually realistic to humans (e.g., see Fig. \ref{fig:simims}), it is hard to ensure that they are perfectly accurate and representative of the true lensing population. 
Additionally, ML models can tend to pick up on subtle biases or artefacts present in the simulations rather than the true astrophysical features that characterise strong lenses. As a result, an ML model that can detect simulated strong lenses with high accuracy does not necessarily perform well at finding lenses in real data. This problem has been well-reported in ML strong lens finders:
\citet{Jacobs17} trained convolutional neural networks (CNNs) to find lenses in CFHTLS data using two different lens simulation approaches, and found that while the CNNs excelled at detecting their respective types of simulated lenses, they underperformed on the other type of simulated lenses and the performance did not translate well to real data;
in \citet{metcalf19}, ten different lens-finding approaches were tested on simulated ground-based data as well as real images from KiDS data, and all ten of them performed significantly better on the simulated data than on the KiDS data; 
in \citet{Pearce-Casey24}, it was found that training the same ML model on different simulated data produced very different performance, and that these networks consistently underperformed on the real data compared to the simulations they were trained on. 
\citet{canameras24} explored how varying training data (both positives and negative) impacts the ability of ML models to recover real lenses from HSC data. As well as finding that the choice of simulated lenses used in training makes a major impact on performance, they found that boosting the fraction of usual contaminants in the negative training class leads to improved performance. 

Although most ML models have relied on simulations, training on real data alone is not without precedent: 
neural networks have been successfully developed to find strong lenses in DESI Legacy Imaging Surveys, using a training set of real lens (of which there were as few as 700) and non-lens images alone \citep{desihuang2020,desihuang2021, desistorfer2024}. Additionally, the limited number of real lenses available for ML training can be mitigated through the use human-in-the-loop pipeline to dynamically build the training set (e.g. Xu et al. in prep.).
However, when training on real lenses alone, it is hard to guarantee the these training sets are representative of the real-world distribution of lenses, and training on these alone may introduce selection biases. Therefore, despite the the issue of the domain gap, training on simulations remains beneficial.


The discrepancy in performance when shifting between domains is a well-known and pervasive challenge in ML, extending beyond lens finding and astronomy \citep{domaingen, zhang2019}, meaning that domain adaptation has been a major topic in computer vision. Humans intuitively learn to recognise high-level semantic features in images, enabling us to generalise across variations in appearance. In contrast, ML models often rely more heavily on low-level properties such as noise patterns and synthetic artefacts.
Simulations can contain subtle statistical differences that are more recognisable in high-dimensional feature spaces, where small mismatches in low-level image statistics can accumulate and become recognisable to ML networks \citep{domaingen}. 
Deep networks, with their large numbers of parameters, have a tendency to exploit these domain-specific features rather than learn generalisable, domain-invariant properties.
As a result, a model trained on one domain (e.g., simulated lenses) often suffers a significant drop in performance when applied to a different domain (e.g., real \Euclid data, \citealt{domainadapt}). 
\textcolor{black}{Domain adaptation techniques such as domain-adversarial neural networks (DANNs; \citealt{danns}) have been explored to address this domain shift between simulations and real observations for other astronomical tasks, such as classifying ionised nebulae \citep{dann_neb} and the study of galaxy mergers \citep{dann_merg}. 
Using pre-trained models has also been shown to enhance adaptation to out-of-domain data \citep{ood2019}, and can improve robustness by combining knowledge from multiple domains \citep{niu2020}.
However, producing domain-invariant ML models remains very challenging, and eliminating the performance drop associated with this domain shift is an open problem.}

The Q1 sample of around 500 \Euclid lens candidates now provides the first statistically meaningful data set for testing and improving ML algorithms on real \Euclid data. In particular, it enables us to: (i) quantify the discrepancy between model performance on simulations and real data; (ii) assess how incorporating real lenses, as well as previously misclassified contaminants, into the training set improves model robustness; and (iii) explore how domain adaptation may scale as the number of confirmed lenses grows in future \Euclid releases. 
In this paper, we explore these aspects through varying the training and testing data with a fine-tuned version of the Zoobot ML model, the most efficient lens-finding network in the initial Q1 lens search \citep{Q1-SP053}. 
Closing the domain gap is expected to produce more accurate, reliable lens-finding models, which will be better equipped to discover lenses in future \Euclid data releases.

The paper is structured as follows. Section \ref{sec:data} covers the data we used that were available prior to Q1, as well as the labelled lenses and non-lenses from Q1 that we can now use for training and testing. In Sect. \ref{sec:method}, we describe the ML model used and our approach for using these new data to investigate ML performance. The results are presented in Sect. \ref{sec:results}, where we first report the results from using the Q1 data to quantify performance in comparison to the pre-Q1 data, and then report the impact of using the Q1 data in training. Finally, the conclusions are presented in Sect. \ref{sec:conclusions}.

\section{Data used for training and testing}\label{sec:data}

The data used in this work consist of the data that were available prior to Q1, which were used to develop the ML models used in Q1, and the labelled Q1 lenses and non-lenses, which resulted from the SLDE search. The former data set (`pre-Q1 lenses and non-lenses', Sect. \ref{ssec:pre-q1-data}) benefits from a larger number of positive training examples, whereas the latter (`Q1 lenses and non-lenses', Sect. \ref{ssec:q1-data}) offers the advantage of being drawn directly from real observations. The data sets are overviewed in Table \ref{tab:data}. Most of these data are introduced in \citet{Q1-SP053}, which includes details on how the distribution of parameters such as Einstein radius varies between the simulations, Q1 lenses, and forecasts. 

\subsection{Pre-Q1 data} \label{ssec:pre-q1-data}
The positive and negative training data that were available prior to Q1 and used originally to train the ML models are detailed in \citet{Q1-SP053}, but key features are outlined here.
The pre-Q1 positives consisted of simulated lenses alone, since the number of known \Euclid lenses at the time were too few.
There were two sets of simulations used. The first simulation set (S1) is from \citet{Q1-SP052}, where the simulations are made by painting lensed arcs onto high-velocity-dispersion luminous red galaxies (LRGs), selected using DESI data. A singular isothermal ellipsoid (SIE) mass model was adopted, with parameters derived from the Sérsic fit, and Einstein radius calculated using the velocity dispersion of the lens and redshifts of the lens and source. Background sources were drawn from the HST Advanced Camera for Surveys F814W high-resolution catalogue \citep{Leauthaud2007,scoville2007,Koekemoer2007}, combined with HSC colour information \citep{Canameras20}. Source images were lensed based on the mass model, downsampled to match \Euclid~{VIS \IE} image resolution, and finally convolved with the telescope PSF after adding the lensed features to the lens images.

The second set of simulations (S2; Metcalf et al. in prep.) was created by selecting \Euclid images of all observed galaxies with \IE < 22, and applying additional cuts to remove stars and reduce the number of face-on spirals. 
Each image was then matched to an object in the Flagship simulation \citep{EuclidSkyFlagship} with a nearest-neighbour algorithm in the space of all magnitudes in all four bands, ellipticity, and redshift, when available for the observed galaxy. The parameters of the Flagship galaxy and dark matter halo were then used to construct a mass model for the lens.
A synthetic source, created by combining between one and four Sérsic profiles, was then placed near or in the tangential caustic. The image of the lensed object was then convolved with the local PSF, and Poisson noise was added.

Notably, both simulation sets employed a strategy of adding simulated arcs on top of real \Euclid images of galaxies in order to include as many features of \Euclid imaging as possible, a well-established strategy of producing the most realistic mock lenses.
This means that any peculiarity of the simulations that could be learnt by the ML models must reside within the lensed arcs, or are due to the fact that lens galaxies used in simulations are not from the exact same underlying distribution as that of real \Euclid lenses.

Additionally, prior to Q1, a catalogue of human-classified non-lenses was compiled from a visual inspection of high-velocity-dispersion galaxies \citep{Q1-SP052}. This catalogue includes approximately 2300 spiral galaxies, 60 ring galaxies, 250 mergers, and 2700 LRGs.
These objects were originally used as the negative class in training.
Prior to Q1 there was no catalogue of classified non-lenses with the same selection cut as that of the Q1 lens search.
Consequently, the distribution of the pre-Q1 non-lenses differs from that of the non-lenses encountered in Q1. In particular, more common false positives, such as spirals and ring galaxies, are overrepresented in the pre-Q1 set relative to LRGs, compared to their distribution in the Q1 data.

\begin{table}[]
\caption{Overview of the data sets used in this work along with their sample sizes.}
\label{tab:data}
{%
\begin{tabular}{llr}
\hline
\hline
\noalign{\vskip 2pt}
                 & Data set                    & Size  \\ \hline
Pre-Q1 positives & Simulations S1              & \num{11057} \\
                 & Simulations S2              & 3737  \\ \hline
Pre-Q1 negatives & Classified non-lenses       & 5000  \\ \hline
Q1 positives     & grade A + grade B lenses    & 497   \\ \hline
Q1 negatives     & Randomly selected Q1 images & \num{40000} \\
                 & ML false positives          & \num{78214} \\ \hline
\end{tabular}%
}
\end{table}

\subsection{Q1 data} \label{ssec:q1-data}
The Q1 data correspond to seven days worth of imaging from \Euclid and make up just 0.45\% of the full EWS. We work with visible imaging data from \Euclid's VIS instrument \citep{EuclidSkyVIS}.
This paper builds upon the work from the original Q1 SLDE lens search, from which 497 strong lens candidates were discovered using these data.
Briefly overviewing the original search, the method involved reducing the original catalogue of 30 million objects to \num{1086554} objects by selecting extended sources having \IE$<22.5$, along with additional selection cuts to remove likely stars and artefact. These \num{1086554} objects were scored by five ML models trained using the data outlined in Sect. \ref{ssec:pre-q1-data}, and these scores informed the selection of objects that were visually inspected by citizen scientists.
In total, around \num{115329} objects were visually inspected, including the top \num{20000} ranked Q1 objects according to Zoobot, objects highly ranked by the other ML models, and \num{40000} randomly selected Q1 objects, aiming to be a representative sample of the Q1 population.
Around 7000 objects, considered likely to be lenses according to the citizen science project, were then graded by strong lensing experts. This resulted in a catalogue of 497 strong lens candidates (corresponding to objects classified as grade A or grade B lenses). 
From a visual inspection of the \num{40000} randomly selected Q1 objects and \num{78214} objects selected by the combined ML models, a catalogue of around \num{100000} classified non-lenses was produced, including a large number of common lens contaminants by construction.
This catalogue is outlined in \citet{Q1-SP048}.

\textcolor{black}{
Of the Q1 strong lens candidates, many objects show very clear lensing features with no other astrophysical explanation. However, without spectroscopic information, it is hard to determine for certain if an object is a true strong lens system or not. Unfortunately, such data are expensive to obtain, and in absence of these ancillary data we rely on visual inspection by strong lensing experts as the most robust method to determine what is or is not a strong lens candidate. \citet{rojas23} demonstrated that averaging the grades assigned by six or more experts is a reliable method of identifying strong lenses, and hence we can be fairly confident that the majority of the candidates in the Q1 sample are likely to be true strong lenses. In the future, spectroscopic data for $\num{10000}$ strong lens candidates will be provided by the 4MOST Strong Lensing Spectroscopic Legacy Survey (4SLSLS; \citealt{4slsls}), which will allow more robust ML models to be trained.
}

We note that some \Euclid lenses exist beyond this SLDE Q1 sample, which we exclude for simplicity. \citet{Q1-SP052} discovered 38 grade A and 40 grade B lenses through the visual inspection of \Euclid imaging of high-velocity-dispersion galaxies. There is some overlap between this catalogue and that of \citet{Q1-SP048}, but approximately 22 (28) grade A (B) lenses were not included in the sample of 497 Q1 lenses considered here. However, many of these were excluded from the Q1 search, either because they did not pass the initial selection cut or because they are outside of the Q1 area and therefore lack data processing consistent with the Q1 data, and hence we do not include them. Following our initial Q1 lens search, there have been other searches through the Q1 data for strong lenses (e.g., Ecker et al. in prep.; Xu et al. in prep.), although these catalogues were not finalised at the time of this work and additionally originate from a different parent sample and so are not considered.

\section{Method}\label{sec:method}

%

\subsection{ML approach}\label{ssec:ml-approach}

We explore the impact of changing the training and testing data for ML performance at lens finding. We test this using a fine-tuned version of the Zoobot foundation model as the ML architecture \citep{Walmsley2023}.
The Zoobot foundation model is pre-trained on around 100 million Galaxy Zoo morphologies across data from a range of surveys \citep{galzoo}, and serves as a base model that can then be fine-tuned for more specific tasks, such as detecting strong lenses.
In the original Q1 search this was the best ML approach for finding lenses \citep{Q1-SP053}, although we note that since the Q1 release, other architectures, such as pre-trained vision transformers, have been shown to perform similarly or better than the Zoobot model (Vincken et al. in prep.). While we expect the general trends to be applicable to most ML models, we note that the quantitative improvement is likely to be architecture-dependent.

The fine-tuning procedure closely follows that of \citet{Q1-SP053}. We use the ConvNeXT-Nano version of the architecture with 15.6 million parameters, and fine-tune the last three blocks.
We use the same image preprocessing (including VIS-only images) and fix the hyperparameters to the best values identified during Q1 training to ensure a fair comparison, varying only the training or testing data sets.
We note that further performance improvement can likely be achieved by optimising the hyperparameters for each specific version of the model, but this paper aims to isolate and study general performance trends, rather than optimising for a single best-performing model.

\subsection{Test sets} \label{ssec:test-data}
When evaluating performance on the pre-Q1 data, we reserve 20\% of the simulations (the combined S1 and S2 sets) and pre-Q1 non-lenses as a held-out test set, while the remaining 80\% is used for training (including epoch-level validation).
For performance evaluation on the Q1 data, we construct a dedicated test set designed to provide a realistic estimate of real-world performance. The positive class consists of 20\% of the Q1 lenses (110 total objects), while the negative class consists of 75\% of a randomly selected Q1 non-lens sample (\num{30000} total objects).
\textcolor{black}{The ratio of positives to negatives in the test set is not reflective of the lensing rate in the Universe, and this is taken into account in all reported metrics. We choose to only include randomly selected Q1 non-lenses, and not the ML-selected false positives, in the test set to ensure the negatives are representative of the true distribution of non-lenses in \Euclid data sets. This established test set is also used to evaluate other \Euclid lens-finding algorithms (e.g., Vincken et al. in prep.).}
The additional Q1 non-lenses that were flagged for visual inspection because they received high scores from the ML models, but were ultimately identified as false positives, are incorporated into the training set, where they do not overlap with the randomly selected negatives.

When evaluating on Q1 data, we use the grades assigned by expert visual inspection as the ground truth, using the 497 grade A and grade B Q1 lenses as the positive sample. The Q1 lens search also resulted in a set of 585 grade C lens candidates that exhibit lens-like features, but could not confidently be classified as lenses, which we include in neither the positive nor negative set for simplicity.
Although humans are not perfect at recognising lenses, especially those of lower signal-to-noise ratio (S/N) or smaller Einstein radii \citep{rojas23, Q1-SP048}, humans remain substantially better at recognising lenses than ML algorithms.
\textcolor{black}{This does mean that the definition of ground-truth is different for the simulations versus Q1 lenses: the Q1 lenses are objects that experts recognise as lenses, whereas the simulations are lenses that can be physically modelled by construction. Therefore, performance on simulations versus Q1 lens candidates is not necessarily an apples-to-apples comparison, and this should kept in mind.}
\textcolor{black}{This distinction implies that the observed performance gap is driven not solely by properties of the training images, but also by the inherent label noise in these observed data.}

\subsection{Performance metrics}
ML performance can be well understood using the receiver operating characteristic (ROC) curve. The ROC curve is the true positive rate (\textrm{TPR}; the fraction of all lenses that are classified as lenses, also known as completeness or recall) against false positive rate (\textrm{FPR}; the fraction of all non-lenses that are classified as lenses). In terms of the number of positives ($N_\sfont{P}$), which is split into true positives ($N_\sfont{TP}$) and false positives ($N_\sfont{FP}$), and negatives ($N_\sfont{N}$), which is split into true negatives ($N_\sfont{TN}$) and false negatives ($N_\sfont{FN}$), these are defined as
\begin{align}
    \textrm{TPR} &= \frac{N_\sfont{TP}}{N_\sfont{P}} = \frac{{N_\sfont{TP}}}{{N_\sfont{TP} + N_\sfont{FN}}}~,\\
    \textrm{FPR} &=\frac{N_\sfont{FP}}{N_\sfont{N}} = \frac{N_\sfont{FP}}{N_\sfont{FP}+N_\sfont{TN}}~.
\end{align}
An ideal classifier can reach ${\textrm{TPR}} =1$ at ${\textrm{FPR}} =0$, and hence has an area under the ROC curve (AUC) of one.
\textrm{TPR} and \textrm{FPR} are invariant to class imbalance, since they are the normalised fraction of lenses and non-lenses that fall within a certain threshold range. These can be converted into purity (equivalent to precision) using the negative-to-positive ratio $\left(N_\sfont{N}/N_\sfont{P}\right)$ as 
\begin{equation}
    \textrm{Purity} = \frac{N_\sfont{TP}}{N_\sfont{TP}+N_\sfont{FP}} = 
    \frac{\textrm{TPR}}{\textrm{TPR}+ \left(N_\sfont{N}/N_\sfont{P}\right)~ \textrm{FPR}}~.
\end{equation}
In Q1, 497 strong lenses were found from an initial sample size of \num{1086554}, meaning there were approximately 2200 non-lenses per lens, $N_\sfont{N}/N_\sfont{P}=2200$. Understanding the trade-off between purity and completeness allows us to better understand the number of lenses expected to be discovered through visually inspecting the top-scored images according to an ML model applied to real data.

While AUC quantifies performance integrated across the full range of thresholds, F1 score can be used as a metric that quantifies the balance of purity and completeness at a given threshold, which translates much better into real-world lens-finding returns. It is the harmonic mean of purity and completeness, or equivalently
\begin{equation}
    \textrm{F1 score}  = \frac{2 ~ N_\sfont{TP}}{2 ~ N_\sfont{TP} + N_\sfont{FP} + N_\sfont{FN}}~.
\end{equation}
We report the maximum F1 value achieved over all possible decision thresholds.
 Because it reflects the trade-off between purity and completeness, this optimum typically lies in the threshold regime where both are reasonably high. This is the most relevant regime for strong lens searches, since this is the range in which candidates above this threshold are forwarded for visual inspection.

For each version of the model with different training or testing data, we repeat the full training and testing process 10 times with different random initialisations in order the quantify the variability of the model and the uncertainty in the performance metrics. 
The statistical uncertainty from bootstrapping within a reserved test set is subdominant compared to the variability from different initialisations (see \citealt{Q1-SP059} for more details).

\section{Results}\label{sec:results}

\subsection{Testing data: simulations versus real lenses} \label{ssec:sim-vs-real}

\begin{figure}
    \centering
    \includegraphics[width=1\columnwidth]{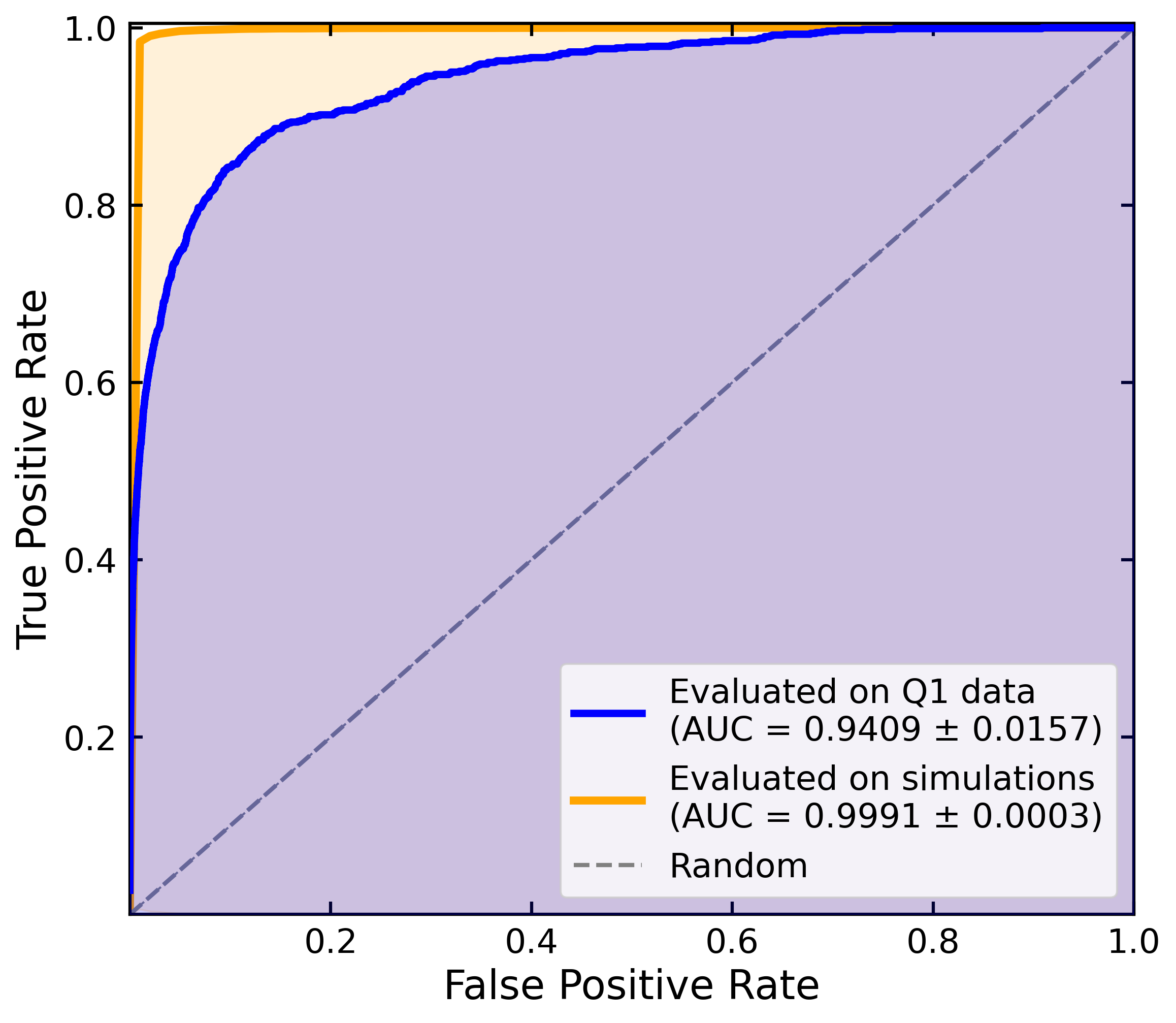}
    \caption{ROC curve of the fine-tuned Zoobot ML model that is trained only on pre-Q1 data (simulated lenses), as evaluated on both pre-Q1 and Q1 data.}
    \label{fig:roc-sims-vs-real}
\end{figure}

We first explore the discrepancy in performance between an ML model evaluated on simulations (in-domain data) versus real lenses (out-of-domain data). 
To do this we train an ML model on a subset of the pre-Q1 data (using simulated lenses only), evaluate it on a separate pre-Q1 test subset from the same parent data set, and then evaluate its generalisation performance on the independent Q1 test set consisting of real lenses (see Sect. \ref{ssec:test-data}).

Figure \ref{fig:roc-sims-vs-real} shows the ROC curve for the same model evaluated on the two different test sets.  When tested on the pre-Q1 data, the model can almost perfectly distinguish the positives from the negatives, with an AUC of $0.9991 \pm 0.0003$. 
In contrast, when tested on the Q1 data, the model performs significantly worse, with an AUC of $0.94 \pm 0.02$. This AUC discrepancy is translated to other metrics, shown in Table \ref{tab:sim-vs-real-metrics}, including maximum F1 score, and purity at set completeness levels. The performance on the in-domain test data is significantly better than the out-of-domain performance by every metric: the same model can recover up to 92\% of the simulated lenses with 100\% purity (zero false-positives), while in the Q1 data the same model recovers 50\% of the lenses with only 24\% purity.

\begin{table}[]
\caption{Performance metrics of the same model evaluated on simulations in comparison to real Q1 data.}
\label{tab:sim-vs-real-metrics}
\resizebox{\columnwidth}{!}{%
\begin{tabular}{llr}
\hline
\hline
                             & \multicolumn{1}{c}{\begin{tabular}[c]{@{}c@{}}Evaluated on \\ simulations\end{tabular}} & \multicolumn{1}{c}{\begin{tabular}[c]{@{}c@{}}Evaluated on \\ Q1 data\end{tabular}} \\ \hline
AUC                          & $0.9991\pm0.0003$                                                                       & $0.9409\pm0.0157$                                                                   \\
F1 score                     & $0.9933 \pm 0.0014$                                                                     & $0.3740 \pm 0.0645$                                                                 \\
Purity at 50\% completeness  & $1.0000 \pm 0.0000$                                                                     & $0.2361 \pm 0.1013$                                                                 \\
Purity at 90\% completeness  & $1.0000 \pm 0.0000$                                                                     & $0.0313 \pm 0.0047$                                                                 \\
Purity at 100\% completeness & $0.9099 \pm 0.0511$                                                                     & $0.0056 \pm 0.0002$                                                                 \\ \hline

\end{tabular}
}
\end{table}


This discrepancy between performance on pre-Q1 and Q1 data indicates that the data distributions are fundamentally different, both for the positive and negative classes, and that the model has implicitly overfitted to features that are not robust across domains.
This highlights a fundamental challenge: simulations and curated training sets, no matter how carefully designed, cannot fully capture the diversity, complexity, and observational nuances of real survey data. As long as the true data distribution remains only partially known, standard metrics such as AUC on in-domain tests give an overly optimistic view of model reliability.
The nature of lens finding creates a circular problem: accurate lens-finding models require a large representative training sample, but obtaining a large representative sample requires accurate lens-finding models. 
Given how realistic current simulations appear, without understanding exactly how they differ from real lenses, it is difficult to make improvements that ensure better transfer of model performance from simulated to real data. Therefore, the only reliable way to close this domain gap is to include real lenses, along with representative non-lenses, in the training sample.

Evaluating performance on the Q1 test set may not perfectly reflect lens-finding abilities, since the Q1 lens sample is not 100\% complete. Most of the Q1 lenses were found due to being scored highly by at least one of five different ML models trained to find lenses in Q1. Therefore, any lenses that are particularly elusive to ML algorithms are likely to have been missed. 
However, the fact that the Q1 lenses presented in \citet{Q1-SP048} originate from a search by five different independently developed ML approaches, along with a visual inspection of randomly selected objects, means that this sample already includes lenses missed by the original Zoobot model and, therefore, the sample should be relatively balanced: it is not the case that the ML network is being tested only on lenses that it previously found.
Even with a slight bias in the sample, we would expect performance on the Q1 test sample to be much more representative of actual performance than that on the simulations.

\begin{figure}
    \centering
    \includegraphics[width=1\linewidth]{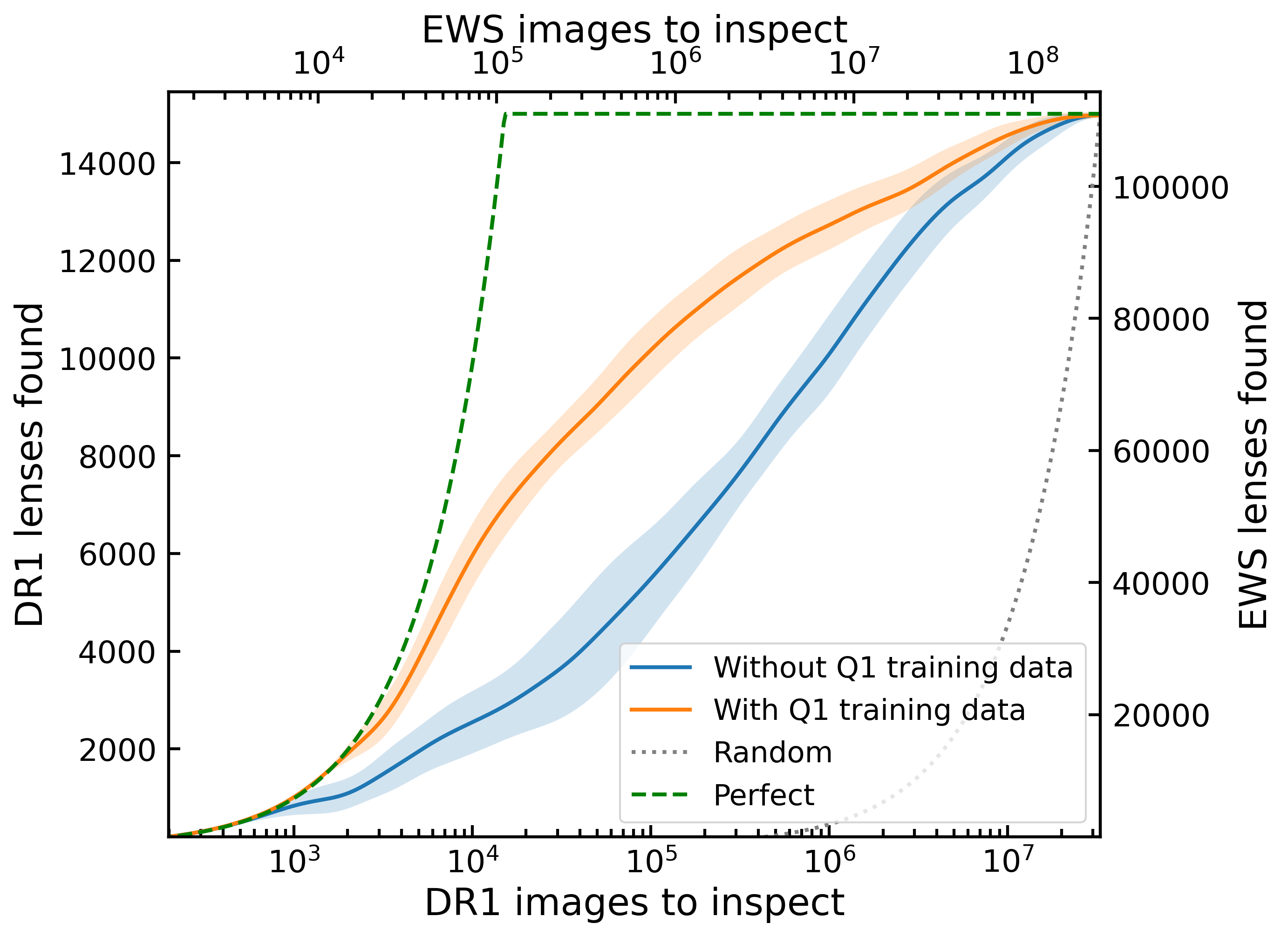}
    \caption{Projected number of lenses discoverable in DR1 and EWS as a function of number of images to inspect, for the network trained with and without the Q1 data.}
    \label{fig:dr1_ews_forecast}
\end{figure}

\subsection{Augmenting pre-Q1 training data with Q1 lenses and non-lenses} \label{ssec:augmenting-training-data}

We next explore the impact of adding the available Q1 data to training. We evaluate the performance on the reserved test set of 20\% of the Q1 lenses and 75\% of the randomly selected images classified as non-lenses.
We can then use the remaining 80\% of the Q1 lenses and randomly selected Q1 images (as well as the objects scored as likely to be lenses but classified as non-lenses) for training. We augment the sample of real lenses by adding four rotations of each lens in order to artificially inflate the number of lenses available for training and to encourage rotational-invariance, but this still only results in 1548 lenses available for training. Given that this sample is still relatively small, we use these Q1 positives and negatives to supplement, rather than replace, the pre-Q1 training data. 

Table \ref{tab:real-world-metrics} shows the same performance metrics as in Table \ref{tab:sim-vs-real-metrics}, but evaluated on the reserved test set of real \Euclid lenses and non-lenses.
In Fig. \ref{fig:dr1_ews_forecast} we use this performance on the reserved test set to extrapolate how many lenses we may find in DR1/EWS as a function of how many images need to be visually inspected. 
To do this, we first compute the TPR (fraction of lenses recovered) as a function of the FPR from the reserved test set of lenses and non-lenses. We then scale these rates to the full survey areas (DR1 and EWS) by extrapolating the total numbers of images and lenses expected, based on the ratio measured in Q1. Finally, by converting the FPR into the corresponding number of images inspected, we obtain the expected number of lenses recovered as a function of images to inspect in each survey.
In the most relevant range for large-scale discovery efforts ($10^5$–$10^6$ inspected images), incorporating Q1 data into the training reduces the number of images that must be inspected by an order of magnitude, while still discovering the same number of lenses.
This translates to a significant increase in the fraction of lenses we could expect to discover in DR1 and EWS using this version of Zoobot alone -- 25\% in the case of DR1 (from visually inspecting \num{500000} images) and 30\% in the case of EWS (from visually inspecting \num{1000000} images). 
Visual inspection of roughly these numbers of images is planned with the help of citizen science through the Space Warps project.

\begin{figure*}
    \centering
    \includegraphics[width=1\textwidth]{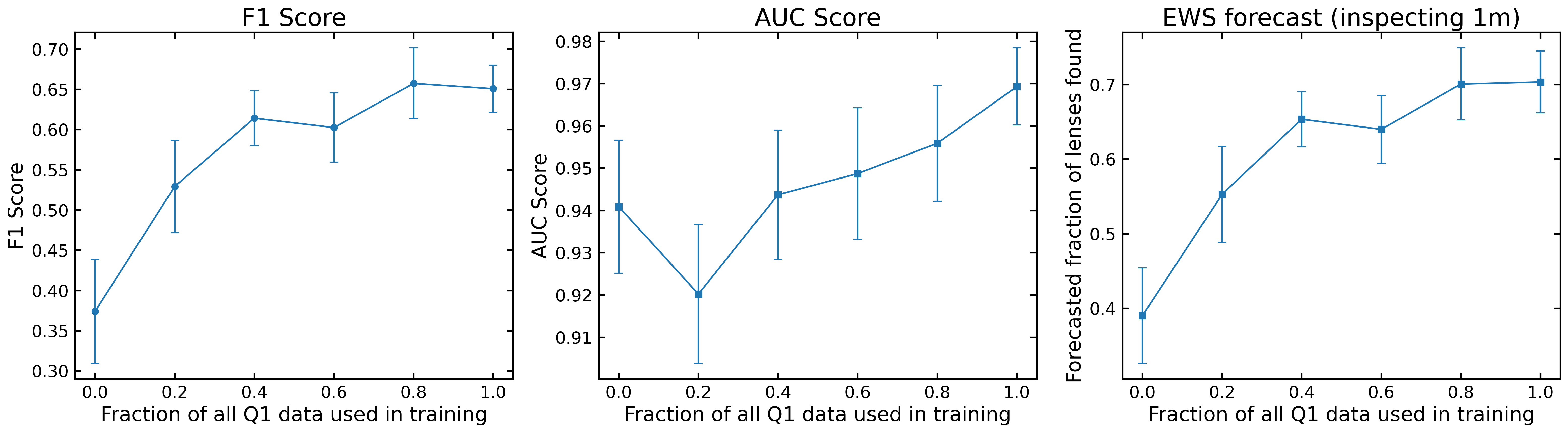}
    \caption{Impact of augmenting the training data with available Q1 lenses and non-lenses, in terms of performance across a range of metrics. These include F1 score and AUC, as well as the projected fraction of lenses that would be discoverable in a data set the size of EWS, assuming that one million images can be visually inspected.}
    \label{fig:adding-Q1-data}
\end{figure*}

To understand the trend in improvement, in Fig. \ref{fig:adding-Q1-data} we show the impact of adding the non-test set Q1 data incrementally to the training data, starting from a model trained only on pre-Q1 positives and negatives. At each increment, another random fraction of the non-test set Q1 lenses are added to the training data. The Q1 non-lenses are added by score, so the first 20\% added to training data are the top 20\% of the Q1 non-lenses that the model initially thought were most likely to be lenses.
This reflects how performance scales with visual inspection depth: the further down the candidate list one inspects, the more labelled non-lenses become available to improve the model. We quantify performance using the F1 score and AUC, which show a trend of improvement as more Q1 data are incorporated into the training set. To relate these metrics to practical outcomes, we use the ROC curves to estimate the fraction of lenses that could realistically be recovered in the EWS by visually inspecting the top \num{1000000} candidates, a ballpark number of images that can feasibly be reviewed.
\begin{figure*}
    \centering
    \includegraphics[width=1\textwidth]{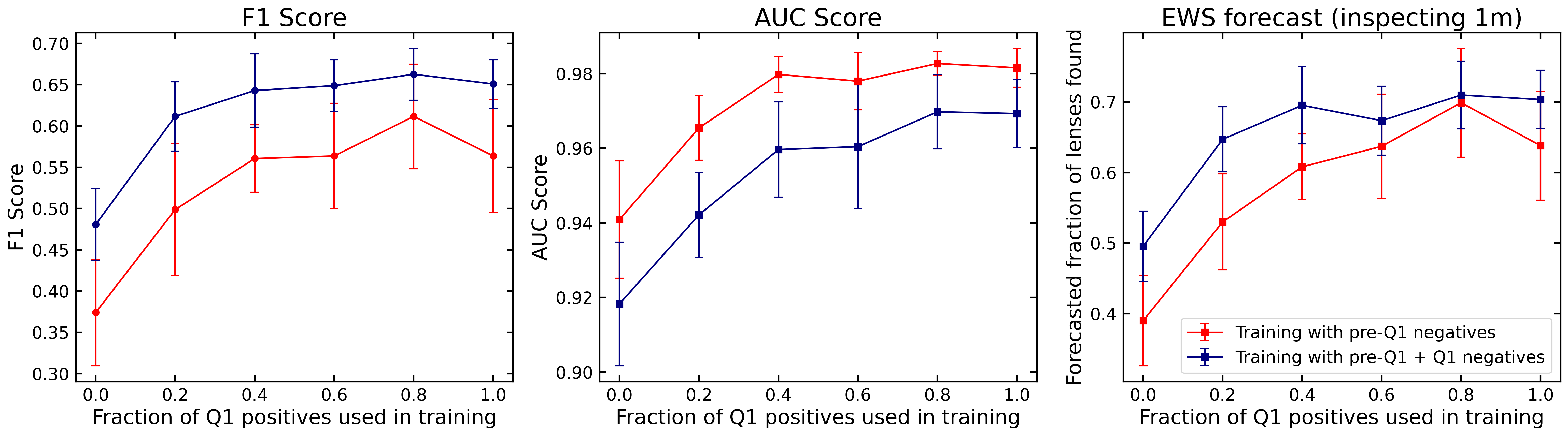}
    \caption{Same as in Fig. \ref{fig:adding-Q1-data}, but isolating the impact of adding Q1 lenses and non-lenses separately. The $x$-axis corresponds to adding the Q1 lenses to the pre-existing training data, and the two lines show the change in performance with and without the addition of the Q1 non-lenses.}
    \label{fig:increasing-tp-only}
\end{figure*}
There is a general upwards trend, though roughly half the improvement is achieved by adding just the first 20\% of the available Q1 data. This suggests that there may be diminishing returns in terms of the performance improvement from adding Q1 data to pre-existing training data. However, given the relatively small size of the Q1 data set, it is hard to confidently extrapolate how much improvement we can expect using images from future \Euclid data releases: DR1 will be around 30 times the size of Q1. 
If performance scales with training data size as a power law, as has been found to be the case in related tasks \citep{scalinglaws}, we would expect substantial improvement once we have a DR1-size data set to train on.
Given the improvement already obtained by re-training on just a few hundred real lenses, it is worthwhile to re-train ML models as soon as new data become available, even if the sample is relatively small.
As more data are acquired, it is expected that further improvements might be achievable by using the Q1 data to replace, rather than augment, the pre-existing training data. This is addressed further in Sect. \ref{sec:q1-only}.

\begin{table}[]
\caption{\textcolor{black}{Same metrics as in Table \ref{tab:sim-vs-real-metrics}, but evaluated on real data rather than simulations. This is shown for a model trained on pre-Q1 data alone (blue curve in Fig. \ref{fig:dr1_ews_forecast}) and the combination of pre-Q1 and Q1 data (orange curve in \ref{fig:dr1_ews_forecast}).}}
\label{tab:real-world-metrics}
\resizebox{\columnwidth}{!}{%
\begin{tabular}{llr}
\hline
\hline
                             & \multicolumn{1}{c}{\begin{tabular}[c]{@{}c@{}}Trained on \\ pre-Q1 data\end{tabular}} & \multicolumn{1}{c}{\begin{tabular}[c]{@{}c@{}}Train on pre-Q1\\ and Q1 data\end{tabular}} \\ \hline
AUC                          & $0.941\pm0.016$                                             & $0.969 \pm 0.009$                                      \\
F1 score                     & $0.374 \pm 0.065$                                           & $0.651 \pm 0.029$                                      \\
Purity at 50\% completeness  & $0.236 \pm 0.101$                                           & $0.838 \pm 0.068$                                      \\
Purity at 90\% completeness  & $0.031 \pm 0.005$                                           & $0.048 \pm 0.036$                                      \\
Purity at 100\% completeness & $0.0056 \pm 0.0002$                                         & $0.0062 \pm 0.0013$                                    \\ \hline
\end{tabular}
}
\end{table}
\subsection{True positives versus false positives}
To understand the discrepancy between performance on the pre-Q1 versus Q1 data, it is informative to investigate how much of the improvement comes from a more representative training class of the positives versus negatives. 
Figure \ref{fig:increasing-tp-only} shows the same metrics as in Fig. \ref{fig:adding-Q1-data}, but plotted as a function of the fraction of the Q1 lenses (rather than all the Q1 data) being added to the training data. This is plotted for two scenarios, one in which all the Q1 non-lenses are also used in the training data, and one in which none of the Q1 non-lenses are added to the training data. This allows us to understand if the problem lies with the simulations not being representative of real lenses, or with the Q1 data containing peculiar non-lens objects that the model has not been trained to recognise.

It can be seen that the majority of the improvement in these metrics is driven by adding the Q1 lenses to the training data: adding the Q1 non-lenses to the training data results in consistently better performance in terms of the number of expected lens discoveries by roughly 5--10\%, but including the Q1 lenses in training increases the fraction of discoverable lenses by roughly 20\%. 
 This suggests that the problem primarily lies with the model misclassifying true lenses as non-lenses. While adding a more diverse set of negatives to the training data can improve purity, obtaining a suitably complete sample of lenses can only be achieved through a robust understanding of the diversity of lens systems.
The pre-Q1 negative set already consisted entirely of pure \Euclid non-lenses, so the domain gap is less of an applicable problem in this case. This explains why the Q1 negatives offer less new information than the positives: the Q1 negative set helps to refine the training data and aids with its understanding of atypical contaminants it might have previously struggled with, whereas the Q1 positives directly expand its knowledge of what genuine lenses look like.


It is interesting to note that according to AUC, using Q1 non-lenses in training is worse than not doing so, despite the fact that according to the other metrics this is not the case. 
\textcolor{black}{Additionally, AUC score is not always well correlated with the number of lenses expected from applying the ML model to real data.}
This highlights a flaw with global metrics like AUC for highly imbalanced classification problems, where only the top-ranked candidates are relevant. As long as a model performs decently well, higher AUC does not necessarily translate to increased lens returns (see Appendix \ref{apdx:A} for a more detailed discussion). In contrast, trends in F1 score are much better correlated with the number of lenses expected to be discoverable in practice.

\subsection{Training on Q1 data alone} \label{sec:q1-only}

\begin{figure}
    \centering
    \includegraphics[width=1\linewidth]{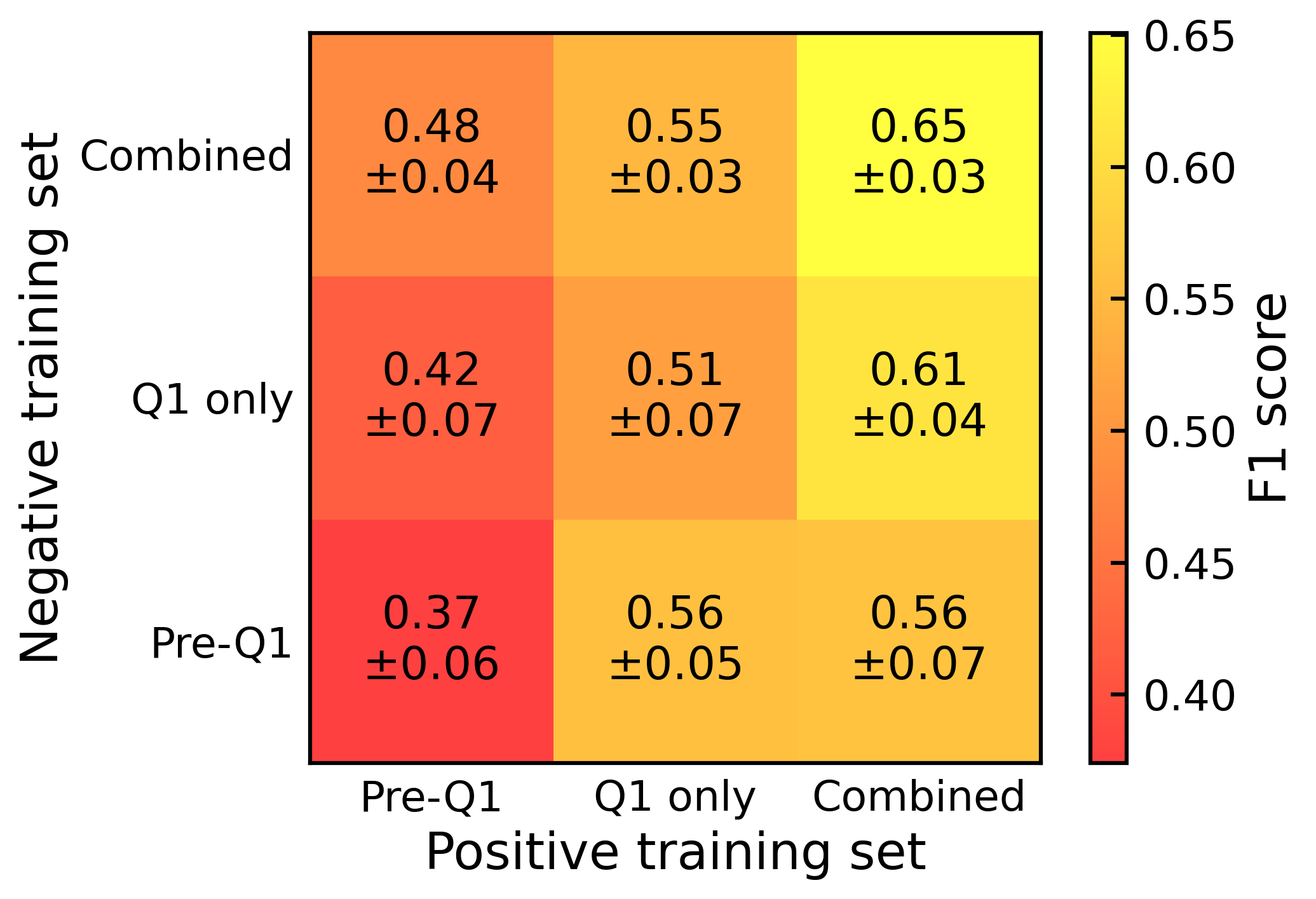}
    \caption{F1 scores achieved by training with different training sets, as evaluated on the reserved Q1 test set.}
    \label{fig:f1-score-heat-map}
\end{figure}

Given the discrepancy between performance on simulations and real data, and the fact that lens-finding neural networks have been successfully trained on lens samples only a few hundred larger than that of Q1, it is interesting to investigate whether a network can be trained on the Q1 data alone. To test this, we vary the composition of both the positive and negative training data independently: each can consist of either the pre-Q1 data only (the simulations and pre-Q1 non-lenses), the Q1 data only, or the combination of both. Figure \ref{fig:f1-score-heat-map} shows how the different combinations of training data impact F1 score, as evaluated on the reserved test set of Q1 lenses and random non-lenses.

Training on the Q1 data alone produces significantly better performance (\mbox{\textrm{F1 score = 0.512 $\pm$ 0.069}}) than training on the pre-Q1 data alone (\mbox{\textrm{F1 score = 0.374 $\pm$ 0.064}}), but does not perform as well as the model trained on the combination of the data (\mbox{\textrm{F1 score = 0.651 $\pm$ 0.029}}). 
Since the F1 score is evaluated only on Q1 lenses and non-lenses, the model trained solely on this subset should perform best, provided that no other limiting factors constrain training, since it is tested on in-domain data.
The superior performance of the model trained on both Q1 and pre-Q1 data suggests that limited training-set size constrains performance when training on Q1 data alone. Once the number of Q1 lenses and non-lenses matches that of the pre-Q1 training size, the model's performance should surpass that of the current combined-data model, since it would then benefit from both sufficient training size and in-domain data.
Alternative training strategies, such as the real-data-focused human-in-the-loop pipeline presented in Xu et al. (in prep.), have proven to be successful at finding lenses without a reliance on simulations. This further reinforces our conclusion that the inclusion of real observational data is critical for maximising the scientific yield of lens-finding campaigns.

In terms of the positive sample, we find that training on the Q1 lenses alone results in a better F1 score compared to training on simulations alone, despite the relatively small sample size of the Q1 lenses.
The fact that the model is being fine-tuned from a pre-trained state, rather than being trained from scratch, likely helps mitigate the usual limitations of smaller training data sets.
The simulations contain complementary information though, since we can see that adding the simulations to the Q1 lenses results in further improvement.
Training on the combination of Q1 lenses and simulations has the added benefit, over training on real lenses alone, that we can be more confident that the simulations cover an appropriate range of lens parameters, such as Einstein radii and S/N. Since many of the Q1 lenses were found by ML models, training on these lenses alone may mean that any selection biases that occur in the ML models will propagate and amplify in a self-reinforcing way. This is of particular concern here, since Q1 is just the start of the \Euclid discoveries: these Q1 lenses will help find lenses in DR1, which in turn will help find lenses in the next data releases. The extent of this potential bias is difficult to quantify due to the limited number of Q1 lenses, but could be investigated further in the future. However, this effect is likely to be mitigated by using multiple ML models, since different approaches generally have different strengths and weaknesses in terms of the types of lenses they find \citep{Q1-SP059, doesmlwork, Nagam25}. 
\textcolor{black}{Additionally, visually inspecting randomly selected \Euclid objects will allow us to calibrate the ML models and allow for lenses to be discovered without this bias.
There is also the argument that the use of the Q1 lenses in training means that the model will be trained to recognise more atypical lens systems that might not have been well accounted for in the simulations. For example, the simulated lenses consisted primarily of lensing by LRGs, while the Q1 sample contained 30--40 late-type disc lenses.}

When considering the negative data, adding the Q1 non-lenses to the training set also leads to an improvement in performance: training on the Q1 only negatives generally results in better performance than training on the pre-Q1 negatives alone, and training on the combination of the two typically results in the best performance. We note that this trend does not hold as well when training on the Q1 positives alone, which could likely be related to the imbalance between the positives and negatives when training in this scenario. However, when training on a sufficiently large positive set (e.g., the combined real and simulated lenses), increasing the negative sample size from $\sim$$10^3$ (pre-Q1 negatives) to $\sim$$10^5$ (Q1 negatives) still improves performance, indicating that the model is fairly robust to class imbalance.
As has been highlighted in previous studies (e.g., \citealt{canameras24}), it is more effective to construct a negative training set that inflates the fraction of common contaminants than to use one that simply mirrors the true distribution of non-lenses in the data. The negative sample obtained from the Q1 lens search contains a large number of typical lens contaminants by construction, so it makes sense that using these for training aids in the model's ability to discern lenses from non-lenses.

\textcolor{black}{
An alternative approach to incorporating real data for mitigating the domain gap is the use of unsupervised domain adaptation techniques, such as DANNs. These architectures include a domain classifier (in addition to the regular label predictor) attached to the feature-extraction layers, which attempts to distinguish between domains (e.g. simulated and real images). Through a gradient reversal layer, the loss from this domain classifier is used to encourage the feature extraction layers to only learn representations that are common across the domains, thereby promoting domain-invariance. This approach has been successfully applied to other astronomical problems (e.g. \citealt{dann_neb}) and could be explored in future work to improve the robustness of strong lens detection models.}


\subsection{Limitations from test-set size}

Throughout this work, performance gains from training on Q1 lenses have been evaluated using a reserved test set consisting of 75\% of the randomly selected Q1 non-lenses (\num{30000} objects) and 20\% of the Q1 lenses (110 objects), leaving sufficient lenses available for training. The set of 110 lenses is a sample size small enough to be influenced by statistical fluctuations. Increasing the proportion of lenses in the test set would reduce the number available for training and thus limit potential performance improvements. Therefore, to assess variability due to the limited test size, we partition the Q1 lenses into five disjoint 20\% subsets (A--E) and alternate which subset is used for testing, with the remaining 80\% used for training.

\begin{figure}
    \centering
    \includegraphics[width=1\linewidth]{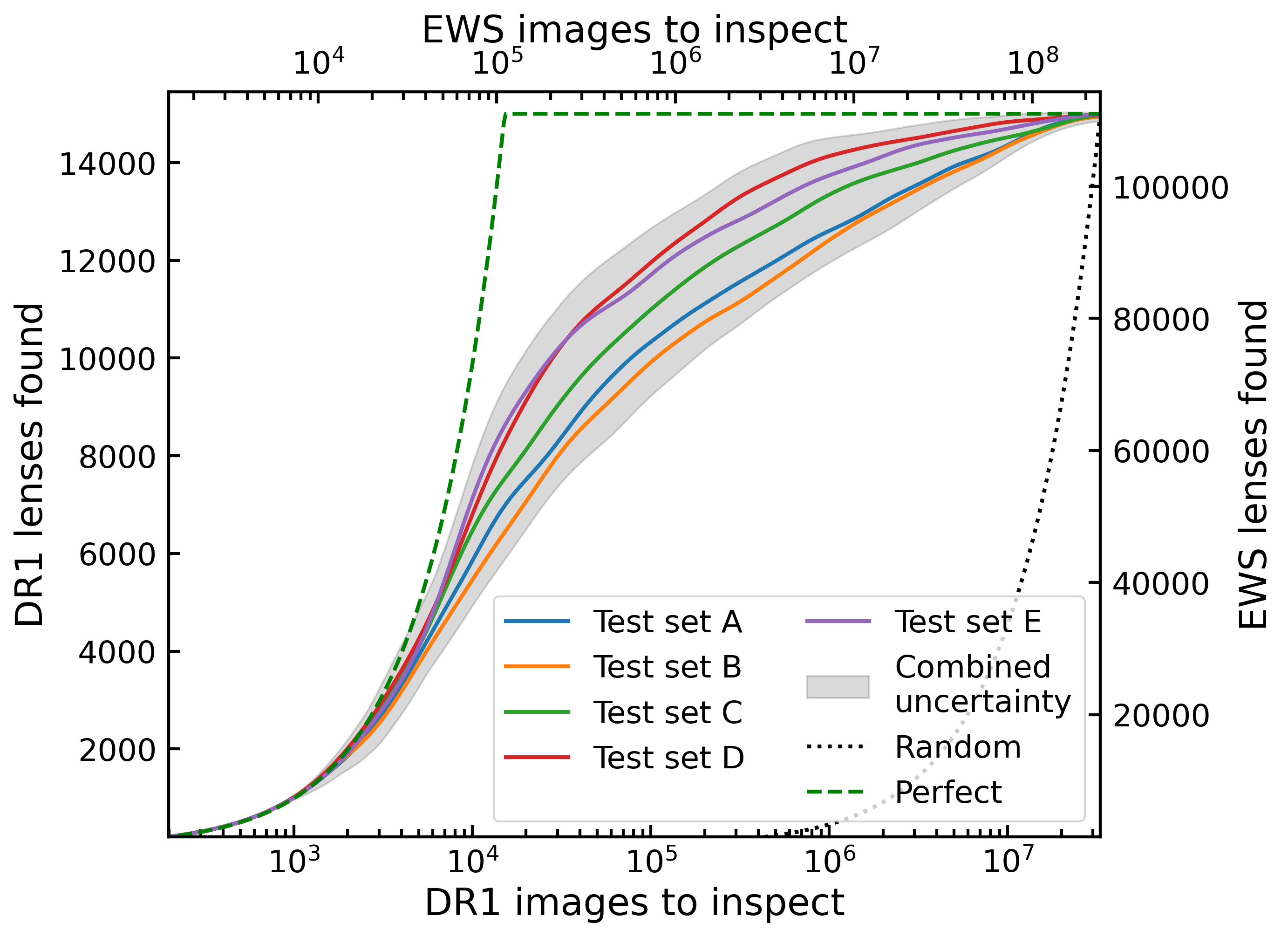}
    \caption{Projected number of lenses discoverable in DR1 and EWS as a function of number of images to inspect (same as Fig. \ref{fig:dr1_ews_forecast}), as evaluated on different 20\% subsets of the Q1 lenses.}
    \label{fig:changing_test_sets}
\end{figure}

Figure \ref{fig:changing_test_sets} displays the variation in performance from varying which Q1 lenses are used for training versus testing, for the version that is trained on the combination of pre-Q1 and Q1 lenses and non-lenses (the best-performing version). Test set A corresponds to the test set that has been used throughout this paper, and its performance is represented by the blue line in Fig. \ref{fig:dr1_ews_forecast}. The variation in performance evaluated on different Q1 lenses is larger than the variation from random initialisations of the model, demonstrating that the limited size of the lens test set is a dominant source of uncertainty in the reported performance. Notably, the performance on test set A is lower than average, suggesting that the performance may have been under-reported, since this subset of the Q1 lenses is harder for the ML model to recognise compared to the broader Q1 lens population.

While the Q1 lens sample size is large enough to understand how lens-finding performance can scale in the future, its limited size means that performance estimates remain sensitive to statistical fluctuations and to the specific lenses included in the test set.
The much larger sample of \num{10000}–\num{15000} lenses expected from \Euclid DR1 will overcome these constraints, reducing uncertainties and allowing us to more confidently extrapolate performance.

\subsection{\textcolor{black}{Exploring the embedding space}}
\textcolor{black}{
While machine learning models are famously difficult to interpret, exploring their embedding space can help us understand their inner workings. This can be achieved using the Uniform Manifold Approximation and Projection (UMAP; \citealt{umap}) algorithm, which projects high-dimensional parameter space into lower dimensions for visualisation, while preserving both local and global structure. By applying UMAP to the embeddings extracted from the layer immediately preceding the classification head, we can visualise how the model organises similar inputs and separates different classes.
}

\textcolor{black}{
Figure \ref{fig:umap} shows the UMAP projection of the pre-Q1 finetuned Zoobot model, mapping simulated images, Q1 lenses, and Q1 non-lenses into a shared embedding space. The plot reveals that Q1 lenses occupy an intermediate position between the simulated lenses and non-lenses. At one extreme (bottom left) lie images with high SNR and complete Einstein rings. A large fraction of simulations fall in this region, whereas fewer real lenses exhibit the same properties. Most Q1 lenses contain less complete rings and often fainter arcs, and while some simulations exhibit these characteristics, their proportion is smaller. The idealised nature of the simulations means that the real lenses that lie closest to the simulated lenses are the grade A lenses, while those nearer the non-lenses tend to be grade B with fainter arcs and less complete rings. 
}

\textcolor{black}{
The Q1 non-lenses are more tightly clustered than the simulated and real lenses, with objects such as extended spirals occupying distinct regions. The greater scatter observed in the positive class relative to the negative class helps understand why incorporating real positives provides a larger performance improvement than adding real negatives: the model already has a more robust understanding of the non-lenses.
Interestingly, a compact subset of the Q1 lenses forms an outlier cluster, corresponding to edge-on disk lenses. 
Although these on average lie closer to the negatives than the positives, the fact that the model can efficiently distinguish these suggests that allowing a multi-class output could be beneficial, and is something to be explored in the future.
}

\textcolor{black}{
The UMAP suggests that part of the difference in performance between simulated and real lenses is due to the overly idealised nature of the simulations. Training on exaggerated simulations can have benefits: such examples present clear lens features, encouraging the model to learn unambiguous characteristics and reduces the risk of confusion. Training on less clear lenses might incentivise the model to pick out non-lensing features. For example,  
\citet{Leuzzi-TBD} demonstrated that including a larger fraction of faint lenses in the training sample can increase misclassification. Additionally, by focusing on unambiguous lenses, the model naturally prioritises candidates that are both more likely to be genuine lenses and are often more scientifically valuable. However, this approach has limitations: the model becomes less robust to fainter or atypical lenses, meaning that achieving a representative and complete sample of the lens population is challenging. Incorporating real observations alongside simulations can therefore help to capture the full diversity of lens morphologies and to ensure robust performance across the range of detectable lenses. 
}
\textcolor{black}{
\begin{figure*}
    \centering
\includegraphics[width=1\textwidth]{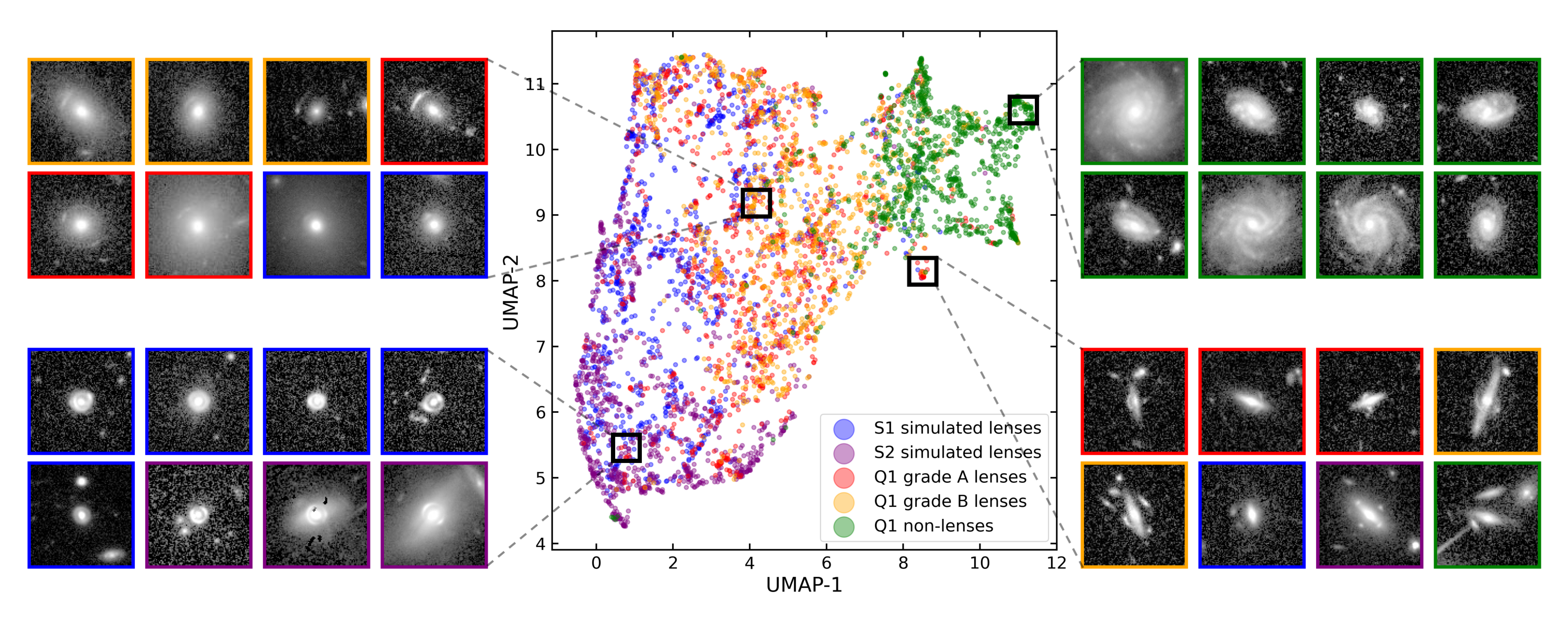}
    \caption{\textcolor{black}{UMAP projection of simulated lenses (S1 and S2), Q1 lenses (grade A and grade B), and Q1 non-lenses in the embedding space of the finetuned Zoobot model. 400 randomly selected points are plotted from each class for clarity.
    Example images shown to the left and right correspond to the highlighted regions of the projection, with coloured borders indicating the associated object class.}}
    \label{fig:umap}
\end{figure*}
}

\section{Discussion and conclusions}\label{sec:conclusions}

The \Euclid mission is set to transform our understanding of cosmology and astrophysics by delivering an unprecedented data set of approximately $\num{100000}$ galaxy-scale strong gravitational lenses, but their discovery within \Euclid’s vast imaging presents a significant challenge. Human visual inspection is infeasible at this scale, making ML models essential for detecting the majority of these lens systems.
However, training robust lens-finding models is challenging. The number of known strong lenses available for supervised learning is still limited, restricting the feasibility of purely real-data-driven approaches.
Consequently, current methods rely heavily on large, diverse sets of simulated lenses, which provide statistical power but inevitably fail to capture the full complexity of real astronomical observations. It is well established that models trained exclusively on one data domain do not always transfer well to another domain, leading to the simulation-to-reality performance gap.

In this work, we have addressed this gap directly by using the 497 strong lenses discovered from the initial search of the \Euclid Q1 data release \citep{Q1-SP048}. These lenses provide the first statistically meaningful sample of real \Euclid lenses, enabling for the first time a direct and quantitative comparison of how models trained on simulations perform on actual \Euclid data.
Furthermore, we investigate strategies to close this gap by combining simulated and real data to improve both completeness and purity in lens discovery.
Our key findings are as follows.
\begin{enumerate}
    \item We confirm a significant discrepancy between model performance on simulated versus real data. A model trained exclusively on simulations that achieves over 90\% completeness with nearly 100\% purity on a simulated test set, only recovers 50\% of real Q1 lenses at a purity of 24\%. This underscores the limitations of relying solely on simulation-based metrics as indicators of real-world performance.

    \item Augmenting the simulation-based training set with real Q1 lenses and non-lenses provides a substantial performance boost. This hybrid training approach can decrease the number of images required to be inspected in future \Euclid data releases by a factor of ten, increasing the projected number of discoverable lenses by 25--30\% for upcoming searches in \Euclid's Data Release 1 and the full EWS. 

    \item The inclusion of real lenses is the primary driver of this improvement, teaching the model the complex, high-fidelity features of observed systems that simulations fail to capture perfectly. Including real non-lenses offers a secondary benefit by helping the model reject common false positives, thereby increasing the purity of the final candidate list. 
    \textcolor{black}{
    The limited number of real lenses available for test sets is a major source of uncertainty in the reported performance of the ML models, but we expect this to be overcome with the growing number of lenses expected from upcoming \Euclid data releases.}
    There will likely be further increases in completeness/purity as the next visual inspection campaigns produce more lenses that can be used for training.

    \item The optimal training strategy is a hybrid approach that combines the statistical power and diversity of large simulation sets with the fidelity of a smaller but growing sample of real lenses. While training on Q1 data alone outperforms training on simulations alone, the limited size of the current real-lens sample means that the combined data set yields the best results. This may change in the future once more \Euclid data have been acquired.
\end{enumerate}

\textcolor{black}{
While the hybrid training approach results in immediate improvements, we must be wary of the risk of a self-reinforcing feedback loop. We are currently at the start of \Euclid lens finding, with the 497 candidates from Q1 represent the seed population for future training sets. As we scale from these hundreds of candidates to the number of lenses expected in the full survey, we must ensure that we understand how our training choices affect the selection function: any biases in the population of lenses that we discover can directly impact the scientific inferences we make from such a sample, particularly if these biases are not quantified and accounted for.
}

\textcolor{black}{
This risk arises from several intersecting factors. Supervised ML models are optimised to recognise objects similar to ones that they are trained on, which in turn are bounded by our knowledge of what we are looking for. While simulations provide a necessary starting point, they rarely capture the full diversity of the real data. For instance, our pre-Q1 simulations did not explicitly prioritise edge-on disk lenses, yet these emerged as a notable population in the real Q1 data. If we strictly optimise our models to find only what we are expecting to find, we risk missing the unexpected discoveries that are an exciting element of the mission.
Beyond the training data, the algorithms themselves possess intrinsic selectivities. For example, the design of CNNs means that they are more adept at learning local spatial correlations, and hence may be better at learning to detect continuous features, such as lensed arcs, compared to Einstein crosses. Another source of selection bias is introduced by training on images recognised as lenses through human visual inspection. If experts consistently overlook certain types of lenses, the ML models will learn to treat these potentially legitimate lenses as negatives.
}

\textcolor{black}{
Ideally, we would detect every lens in the survey with 100\% completeness and purity, resulting in a catalogue perfectly representative of the strong lensing distribution in the Universe. Since this is infeasible, understanding the completeness of our detection rates as a function of the lens parameter space is important. Several strong lensing science cases can be enabled by obtaining a subset of the lensing population with 100\% completeness, at the expense of a smaller sample size \citep{ale22,zhou24}.
To ensure a less biased catalogue and robustly characterise the selection function, there are a few potential strategies.
First, we must continue the visual inspection of random images to identify lens candidates without the ML bias. While this method is still subject to human bias, it is free from the specific morphological biases of the ML models. 
Future spectroscopic surveys, such as 4MOST, will be able to confirm many of these lens candidates, thereby providing a more reliable and consistent definition of what constitutes a strong gravitational lens.
Second, we should employ multiple independent ML approaches, as different approaches will inherently prioritise different regions of the parameter space. For example, the pipeline developed by Xu et al. (in prep.) minimises reliance on simulations in favour of iterative training on human-identified real lenses. In their application to Q1, this approach uncovered 91 previously missed grade B candidates and four new grade A candidates. This prioritising of grade B candidates illustrates the specific utility of their training strategy: while it carries a stronger imprint of human selection bias, it is effective at recovering more ambiguous candidates that constitute a large range of the lens candidate population and are more likely missed by models trained on idealised simulations.
Finally, we must quantify the domain gap in the selection function. While we can easily measure selection functions en masse using simulations, we must determine if these measurements are transferable to real data. A first step is to explore whether the recoverability trends observed in simulations (e.g., improved detection at high SNR) translate equivalently to real lenses.
Combining these diverse discovery channels and using simulated lenses to map their selection functions will ensure the final \Euclid lens catalogue is both vast and scientifically valuable. 
}

\textcolor{black}{
With these selection effects quantified, we can confidently scale up the search by iteratively updating our models as new data releases arrive. This approach optimises the use of visual inspection time, balancing the speed of ML with the reliability of human visual inspection, and offers the best path to build the largest sample of strong gravitational lenses to date.}

\section{Data availability}
This paper makes use of the Euclid Quick Release 1 data \citep{Q1cite}, covered by \citet{Q1-TP001}.

\begin{acknowledgements}

NEPL is supported through a graduate studentship from the UKRI STFC and the University of Portsmouth.

Numerical computations were carried out on the SCIAMA High Performance Compute (HPC) cluster, which is supported by the ICG and the University of Portsmouth.

This work has received funding from the European Research Council (ERC) under the European Union's Horizon 2020 research and innovation programme (LensEra: grant agreement No 945536). TEC is funded by the Royal Society through a University Research Fellowship.

SS has received funding from the European Union’s Horizon 2022 research and innovation programme under the Marie Skłodowska-Curie grant agreement No. 101105167 -- FASTIDIoUS.

\AckEC

\AckQone

\end{acknowledgements}

%
%

\bibliography{Euclid, Q1, bib}

%
%

\begin{appendix}
  \onecolumn 
  
\section{Translating ROC curves into expected lens discoveries\label{apdx:A}}
\begin{figure}
    \centering
    \begin{minipage}[b]{0.48\textwidth}
        \centering
        \includegraphics[width=\textwidth]{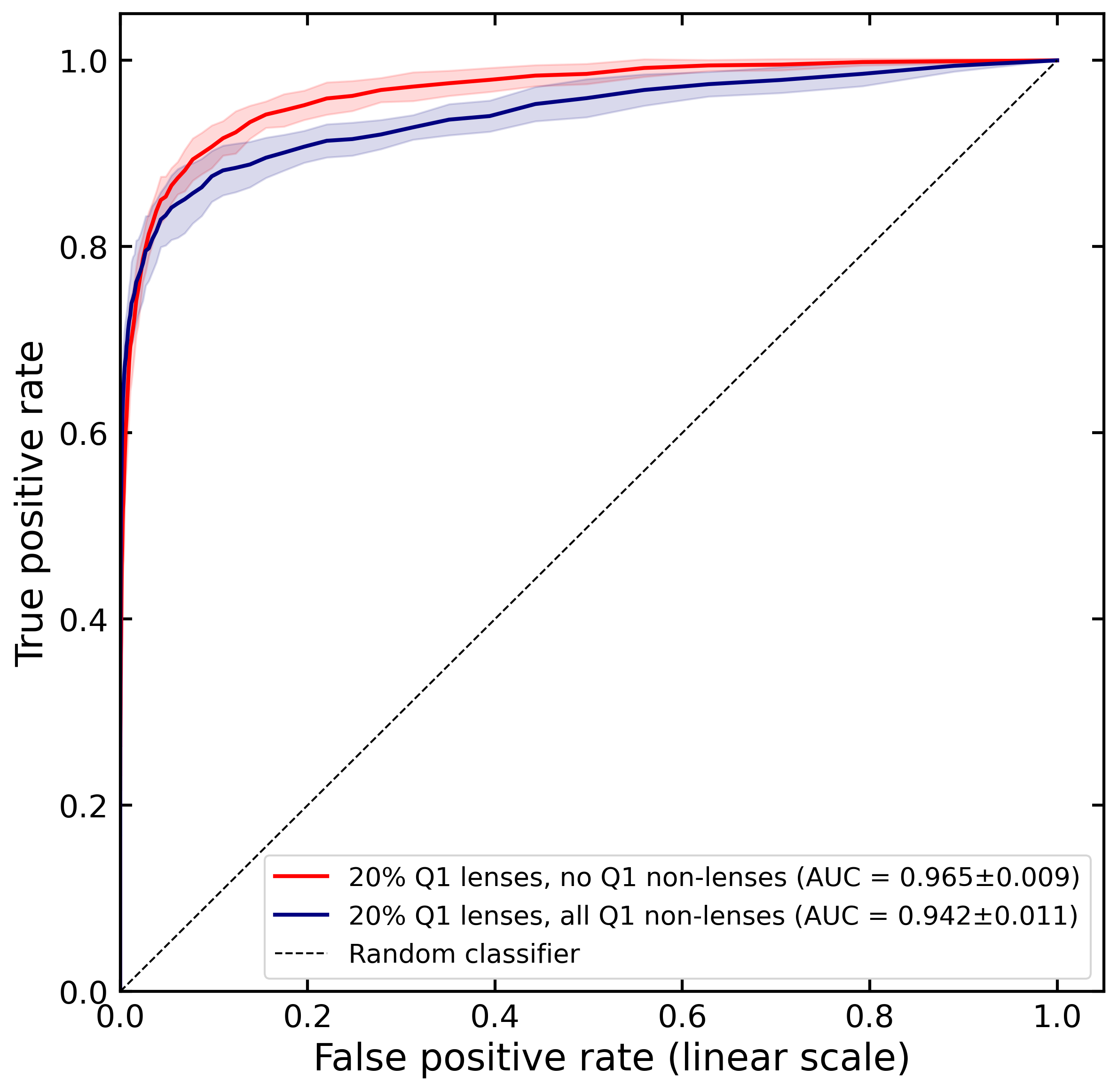}
    \end{minipage}
    \begin{minipage}[b]{0.49\textwidth}
        \centering
        \includegraphics[width=\textwidth]{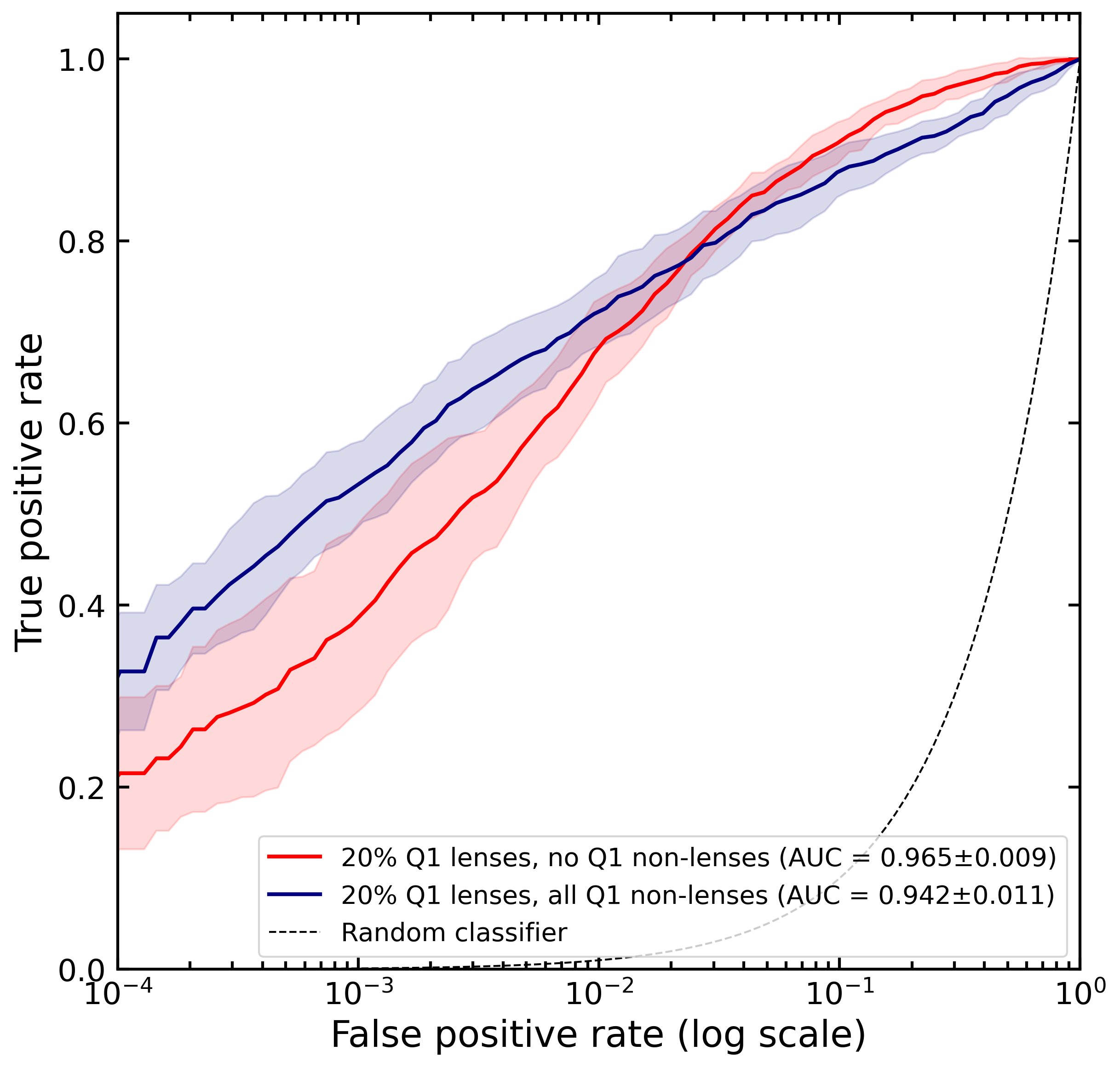}
    \end{minipage}
    \caption{ROC curves \textcolor{black}{in linear scale (\textit{left}) and log scale (\textit{right})} showing the performance of two different versions of the Zoobot network, each with different augmentations to the training data: one using 20\% of the Q1 lenses and none of the Q1 non-lenses; and the other using 20\% of the Q1 lenses and all of the Q1 non-lenses.}
    \label{fig:roc-comparison}
\end{figure}

Figure \ref{fig:roc-comparison} shows two ROC curves for models with two different training data sets. The first augments the pre-Q1 training data with 20\% of the Q1 lenses and none of the Q1 non-lenses, and the second with 20\% of the Q1 lenses and all of Q1 non-lenses.
The former has a better AUC ($0.965\pm0.009$) than the latter ($0.942\pm0.011$). However, the version that includes the Q1 non-lenses in the training data outperforms the other version in the range $\textrm{TPR}<0.8$. Despite this range representing a very small fraction of the ROC curve, it is the range of interest for lens finding. Even visually inspecting within the range $\textrm{FPR}<0.1$ means that 10\% of all the negatives have to be visually inspected: this translates to around 3 million images in a DR1-size sample. The discrepancy in the AUC comes from the difference in ability to recover lenses in the range $0.1<\textrm{FPR}<0.6$, a range that lies beyond the scope of what could be visually inspected. For this reason, a higher AUC does not directly correlate with better lens-finding performance.

\end{appendix}

\label{LastPage}
\end{document}